\documentclass[fleqn,usenatbib]{mnras}

\usepackage{newtxtext,newtxmath}

\usepackage[T1]{fontenc}
\usepackage{multirow}
\usepackage{booktabs}
\usepackage{orcidlink}
\usepackage{verbatim}

\DeclareRobustCommand{\VAN}[3]{#2}
\let\VANthebibliography\thebibliography
\def\thebibliography{\DeclareRobustCommand{\VAN}[3]{##3}\VANthebibliography}

\usepackage{graphicx}   
\usepackage{amsmath}    
\usepackage{subfigure}  
\usepackage{siunitx}
\usepackage[figuresleft]{rotfloat}
\usepackage{CJKutf8}
\usepackage{appendix}

\newcommand{\SigSFR}{$\mathrm{\Sigma_{\mathrm{SFR}}}$}
\newcommand{\ha}{H$\mathrm{\alpha}$}
\newcommand{\dphi}{$\mathrm{\Delta\phi}$}
\newcommand{\re}{$R_e$}
\newcommand{\offset}{$\mathrm{\Delta log(O/H)}$}
\newcommand{\dfth}{$\mathrm{D_{4000}}$}

\title[ISM in Spiral Galaxies]{The MAGPI Survey: Effects of Spiral Arms on Different Tracers of the Interstellar Medium and Stellar Populations at $z \sim 0.3$}

\author[Qian-Hui Chen et al.]{
    Qian-Hui Chen (陈千惠)$^{\orcidlink{0000-0002-4382-1090}}$, $ ^{1,2}$\thanks{E-mail: Qianhui.Chen@anu.edu.au}
	Kathryn Grasha$^{\orcidlink{0000-0002-3247-5321}}$, $ ^{1,2,14}$\thanks{ARC DECRA Fellow}
	Andrew J. Battisti$^{\orcidlink{0000-0003-4569-2285}}$, $^{1,2}$
	Emily Wisnioski$^{\orcidlink{0000-0003-1657-7878}}$, $^{1,2}$
	\newauthor
	Trevor Mendel$^{\orcidlink{0000-0002-6327-9147}}$, $^{1,2}$
        Piyush Sharda$^{\orcidlink{0000-0003-3347-7094}}$, $^{3}$
        Giulia Santucci$^{\orcidlink{0000-0003-3283-4686}}$, $^{4,2}$
        Zefeng Li (李泽峰)$^{\orcidlink{0000-0001-7373-3115}}$, $^{1,2}$
        Caroline Foster$^{\orcidlink{0000-0003-0247-1204}}$, $^{5,2}$
 \newauthor
        Marcie Mun$^{\orcidlink{0000-0002-3706-9955}}$, $^{1,2}$
	Hye-Jin Park$^{\orcidlink{0000-0002-9809-6631}}$, $^{1,2}$
	Takafumi Tsukui$^{\orcidlink{0000-0002-1499-6377}}$, $^{1,2}$
        Gauri Sharma$^{\orcidlink{0000-0002-6070-2851}}$, $^{6,7}$
        \newauthor
        Claudia D.P. Lagos$^{\orcidlink{0000-0003-3021-8564}}$, $^{4,8}$
        Stefania Barsanti$^{\orcidlink{0000-0002-9332-5386}}$, $^{1,2,9}$
        Lucas M. Valenzuela$^{\orcidlink{0000-0002-7972-9675}}$, $^{10}$
        Anshu Gupta$^{\orcidlink{0000-0002-8984-3666}}$, $^{11, 2}$
        \newauthor
        Sabine Thater$^{\orcidlink{0000-0003-1820-2041}}$, $^{12}$
        Yifei Jin (金刈非)$^{\orcidlink{0000-0003-0401-3688} 13,1,2}$
        and Lisa Kewley$^{\orcidlink{0000-0001-8152-3943} 13,1,2}$
	\\
	$^{1}$Research School of Astronomy and Astrophysics, Australian National University, Canberra, ACT 2611, Australia\\
	$^{2}$ARC Centre of Excellence for All Sky Astrophysics in 3 Dimensions (ASTRO 3D), Australia\\
        $^{3}$Leiden Observatory, Leiden University, P.O. Box 9513, NL-2300 RA Leiden, The Netherlands\\
        $^{4}$International Centre for Radio Astronomy Research, The University of Western Australia, 35 Stirling Highway, Crawley, WA 6009, Australia\\
        $^{5}$School of Physics, University of New South Wales, Sydney, NSW 2052, Australia\\
        $^{6}$University of Western Cape, South Africa\\
        $^{7}$INFN-Sezione di Trieste, via Valerio 2, I-34127 Trieste, Italy\\
        $^{8}$Cosmic Dawn Center (DAWN)\\
        $^{9}$Sydney Institute for Astronomy (SIfA), School of Physics, The University of Sydney, NSW 2006, Australia\\
	$^{10}$Universitäts-Sternwarte, Fakultät für Physik, Ludwig-Maximilians-Universität München, Scheinerstr. 1, 81679 München, Germany\\
        $^{11}$International Centre for Radio Astronomy Research (ICRAR), Curtin University, Bentley WA 6102, Australia\\
        $^{12}$Department of Astrophysics, University of Vienna, Türkenschanzstraße 17, 1180 Vienna, Austria\\
        $^{13}$Institute for Theory and Computation, Harvard-Smithsonian Center for Astrophysics, Cambridge, MA 02138, USA \\
    $^{14}$Visiting Fellow, Harvard-Smithsonian Center for Astrophysics, 60 Garden Street, Cambridge, MA 02138, USA\\
}

\date{Accepted XXX. Received YYY; in original form ZZZ}

\pubyear{2023}

\begin{document}
\begin{CJK}{UTF8}{gbsn}
	\label{firstpage}
	\pagerange{\pageref{firstpage}--\pageref{lastpage}}
	\maketitle
	
	\begin{abstract}
            Spiral structures are important drivers of the secular evolution of disc galaxies, however, the origin of spiral arms and their effects on the development of galaxies remain mysterious. 
            In this work, we present two three-armed spiral galaxies at $z \sim 0.3$ in the Middle Age Galaxy Properties with Integral Field Spectroscopy (MAGPI) survey. 
            Taking advantage of the high spatial resolution ($\sim \ 0.6\ ''$) of the Multi-Unit Spectroscopic Unit (MUSE), we investigate the two-dimensional distributions of different spectral parameters: \ha, gas-phase metallicity, and \dfth. 
            We notice significant offsets in \ha\ ($\sim$ 0.2 dex) as well as gas-phase metallicities ($\sim$ 0.05 dex) among the spiral arms, downstream and upstream of MAGPI1202197197 (SG1202). This observational signature suggests the spiral structure in SG1202 is consistent with arising from density wave theory. 
            No azimuthal variation in \ha\ or gas-phase metallicities is observed in MAGPI1204198199 (SG1204), which can be attributed to the tighter spiral arms in SG1204 than SG1202, coming with stronger mixing effects in the disc.
            The absence of azimuthal \dfth\ variation in both galaxies suggests the stars at different ages are well-mixed between the spiral arms and distributed around the disc regions. 
            The different azimuthal distributions in \ha\ and \dfth\ highlight the importance of time scales traced by various spectral parameters when studying 2D distributions in spiral galaxies. 
            This work demonstrates the feasibility of constraining spiral structures by tracing interstellar medium (ISM) and stellar population at $z\sim 0.3$, with a plan to expand the study to the full MAGPI survey. 
            
	\end{abstract}
	
	
	
	
	\section{Introduction}\label{sec:intro}
     In the local Universe, the majority of star formation takes place in spiral galaxies \citep{Brinchmann_2004} and up to two-thirds of all massive galaxies are spiral galaxies \citep[][]{Lintott_2011, Willett_2013}.
     The interplay between star formation activity and the influence of spiral structures is closely related to 1) the physical conditions and distribution of the gas in the interstellar medium (ISM), and 2) the processes by which stars are created from the molecular gas. 
     Spiral structures are complex and diverse, which is demonstrated by the variety in their number of arms, pitch angles, amplitudes and longevity \citep{Dobbs_2014}.
     Despite their ubiquitous nature, we know surprisingly little of the physical origin of spiral features and how they evolve over cosmic time \citep{Sellwood_2022}.

	Previous studies investigating spiral-driven evolution have revealed the physical complexities of the origin of spiral features, including self-excited instabilities in equilibrium discs and gravitational instabilities due to interactions with companions, halos, and stellar bars. We summarise the two main physical mechanisms for the origin of spiral arms in disc galaxies:
	\begin{itemize}
	\item (Quasi-stationary) density wave theory \citep[][]{Lin_1966}: 
     spiral arms are density waves that evolve slowly and rotate at a fixed angular velocity (i.e., pattern speed) across the galaxy. The stars and gas in the disc orbit at different angular velocities, depending on their distances to the galaxy centre, which is called differential rotation.  
     The co-rotation radius is the radius where the angular velocity of the material matches the pattern speed \citep[e.g.,][]{Pour-Imani_2016, Peterken_2019}. 
     In this theory, the material is left behind the density wave outside the co-rotation radius but overtakes the density wave while inside the co-rotation radius.
     When entering the spiral potential, the gas will collapse \citep[][]{Toomre_1977} and undergo shocks \citep[][]{Roberts_1970,Bonnell_2006}, which are observable as periodic episodes of intense and instantaneous star formation \citep{Aouad_2020}.
     The formation of a global spiral pattern from the density wave theory is caused by an instability of the disc as a direct result of self-gravity \citep{Pettitt_2017}. 
     According to the density wave theory, spiral galaxies with more than three spiral arms are less stable compared to grand-design spiral galaxies with two arms \citep{Thomasson_1990}. 
     Simulations report long-lived grand-design spiral galaxies which resemble the density wave theory, although the overall spiral pattern is transient and recurrent \citep{Sellwood_2011,DOnghia_2013, Sellwood_2014}.
     The maintenance (or the necessity of maintenance) of long-lived spiral waves is still an open question. 
     
	\item Dynamic spirals \citep[][]{Sellwood_1984, Fuchs_2001, Dobbs_2014}: 
     in the dynamic material wave model, a leading density pattern transfers into a trailing one due to the shear caused by a differentially rotating disc. 
     During the rotation, the leading mode is amplified into a spiral arm as a result of the self-gravity, through a mechanism called swing amplification \citep{Goldreich_1965}. 
     In this mechanism, spiral arms are simply a superposition of many unstable waves \citep{Michikoshi_2016}. There is no specific pattern speed for dynamic spirals or significant offset between the rotation of spirals and the disc. 
     Thus, gas does not pass through the spiral arms but instead falls into the minimal potential from both sides of the arms \citep[][]{Dobbs_2008, Wada_2011}. The spirals generated by swing amplification are transient, recurrent, and short-lived \citep{Sellwood_2014,Sellwood_2019}.
     Swing amplification typically assembles flocculent or multi-armed galaxies but is capable of generating grand-design spiral arms \citep[][]{Baba_2015}. \citet{Sellwood_1984} argue for short-lived (a few $\sim$ 100~Myr) dynamic spirals due to the heating from dissipation. On the other hand, \citet{Fujii_2011} and \citet{DOnghia_2013} demonstrate the possibility of long-lasting (up to $\sim$ 10~Gyr) dynamic spirals using highly spatially-resolved N-body simulations.
	\end{itemize}

    The density wave theory and dynamic spiral theory are not exclusive to each other. Instead, the two mechanisms can explain the origin and assembly of different spiral arms.
    Many studies have been conducted to investigate the feasibility of these two theories through simulations. 
	Since the 1970s, transitory spiral features have been seen in some simulations \citep[][]{Miller_1970, Hohl_1971, Hockney_1974}, which gives us a glimpse of the fundamental physics of spirals. Assuming different models of spirals, simulations test and improve the theories by revealing the longevity of spiral arms \citep[][]{Oh_2008,Fujii_2011,Struck_2011,DOnghia_2013, Sellwood_2014}, distributions of ISM properties \citep[][]{Dobbs_2010,Grand_2012a, Grand_2012b, Pettitt_2020}, motions of stars in spiral galaxies \citep[][]{Lindblad_1996,Baba_2013,Ramon-Fox_2022}, etc. 
    These studies find that both the density wave theory and dynamic spirals are viable mechanisms for the formation of spiral arms. 

    Although non-trivial and difficult, there have been many attempts to distinguish the physical mechanism driving the formation of spiral arms from observations in the local Universe.
    The density wave theory adopts a fixed pattern speed across the disc while the dynamic spirals induce the spiral arms to rotate with the disc at a radially decreasing angular velocity. 
    The simplest detection of pattern speed is the co-rotation radius measured by the residual velocity maps of $\mathrm{^{12}C^{16}O}$ and \ion{H}{i} \citep{Elmegreen_1989, Canzian_1993, Sempere_1995}, although this method is highly subject to measurement uncertainties. 
    \citet{Peterken_2019} detect the pattern speed by mapping the ongoing star formation and previously formed young stars as a direct test of the density wave theory.
    Another indirect method to determine the pattern speed is to detect the location of spiral-shaped shock waves (hereafter spiral shocks). 
    The density wave theory predicts spiral shocks to be found on one side of the stellar potential within the co-rotation radius and on the opposite side outside the co-rotation radius.
    This results in an offset whose width is determined by the difference between pattern speed and material speed.
    In the dynamic spiral theory, we do not expect an offset on both sides of the spiral arms, as the material in the disc does not cross the spiral arm potential.
    Therefore, the distributions of gas-phase ISM properties and stellar populations are different depending on the assumed theory, which enables us to interpret the observations and understand the origin of spiral arms in the observed galaxies.
    This method has been applied to several nearby galaxies \citep[e.g., ][]{Egusa_2004, Tamburro_2008, Egusa_2009, Pour-Imani_2016, Abdeen_2022}, but not all observed galaxies are consistent with the prediction of the density wave theory \citep{Foyle_2011}.

    Integral Field Spectrographs (IFS) enable investigation into the radial and azimuthal distributions of stellar and gaseous ISM properties of galaxies. 
    Large IFS surveys provide spatially-resolved distributions of oxygen abundances, kinematics, and stellar populations \citep{Falcon-Barroso_2006, Marino_2012, Rodriguez-Baras_2018, Kreckel_2019, Grasha_2022, Chen_2023}, to constrain the formation and evolution of spiral galaxies. 
    These studies provide unprecedented insight into the dynamics and metal variations of spiral galaxies in the local Universe.
	However, most IFS studies of spiral galaxies are limited to spirals in the local Universe, leaving a gap in our understanding of spiral patterns in high-redshift galaxies, when the very first galaxies settled down to develop spiral features. 
    Even though observations have found the existence of spiral galaxies up to $z\sim 2$ \citep{Law_2012, Margalef-Bentabol_2022}, the formation of these spiral galaxies remains mysterious.
    The formation of spiral arms is assumed to happen in thin stable discs \citep{Toomre_1977, Sellwood_2014} while galaxy discs are thicker, more gas-rich and less stable at higher redshifts \citep{Wisnioski_2015, Crain_2015}. 
    The effects of the ISM become increasingly important in the build-up of spiral arms at higher redshifts \citep{Wada_2011, Ghosh_2016}. 
    Spiral arms are also sensitive to environmental effects, including gas accretion and galaxy mergers \citep{Wada_2011, Martig_2012, Dobbs_2014}.
    To understand the physics of spiral formation and when it dominates the star formation activities in star-forming galaxies, we need to push detailed observations to higher redshifts. 
    Using spatially-resolved IFS data targeting galaxies at a higher redshift, it is possible to discriminate density wave theory/dynamic spirals at an earlier stage of the formation of spiral galaxies. 

    Taking advantage of the adaptive optics of MUSE, the Middle Ages Galaxy Properties with Integral Field Spectroscopy \citep[MAGPI;][]{Foster_2021} survey is carried out to target 60 massive galaxies and their satellites at 0.25 $\textless z \textless$ 0.35 (3-4 Gyr ago). 
    Since the dynamical, morphological, and chemical properties of galaxies undergo significant transformation at $z \sim 0.3$, the MAGPI survey gives an important window in the cosmic timeline and fills up the gap in our understanding of the build-up of spiral features. 
    The spatial resolution of MAGPI, which reaches 0.6 - 0.8 arcsec (full width at half maximum; FWHM), enables us to distinguish the spiral arms and the inter-arms without the use of gravitational lensing.
    In this work, we study the two-dimensional distributions of different ISM properties and stellar population in two spiral galaxies at $z \sim 0.3$ at a high spatial resolution, shedding light on the origin and growth of spiral arms. 
    This paper demonstrates the feasibility of tracing ISM properties and their linkage to stellar spiral structures at $z \sim 0.3$, with a plan to expand the study to the full MAGPI spiral sample upon completion of the full dataset.

    In this paper, we present our first work on exploring spiral galaxies with IFS observations at $z \sim$ 0.3. 
    We introduce our sample in Sec~\ref{sec:data_sample}. 
    We present our analysis of two gas ISM properties as well as stellar age in Sec~\ref{sec:analysis} and the results are then discussed in Sec~\ref{sec:discussion}. 
	Throughout this paper, we adopt the \citet{Chabrier_2003} initial mass function (IMF), and a flat cosmology with $\Omega_{\mathrm{\Lambda}}$ = 0.7, $\Omega_\mathrm{m}$ = 0.3, $H_0$ = 70 km s$^{-1}$ kpc$^{-1}$.

\section{Data and Sample}\label{sec:data_sample}
    \subsection{Data}\label{subsec:data} 
    We use the observational data from the MAGPI\footnote{\url{https://magpisurvey.org}} survey (35 fields at the time of starting this work, out of 56 MAGPI fields). 
    MAGPI is a large program on the European Southern Observatory Very Large Telescope (VLT) which is still in progress at the time of writing (Program ID: 1104.B-0536).  
    The observations are achieved through 56 $\times$ 4.4h on-source exposures of independent fields with Ground Layer Adaptive Optics (GLAO) on VLT/MUSE. The MAGPI sample is also supplemented with two archival MUSE fields of massive clusters at $z\sim0.3$: Abell 370 and Abell 2744.
    
    The science goals, design, and observing technique are described in detail in \citet{Foster_2021}. 
    In brief, the survey aims to reveal and understand the physical processes responsible for the rapid transformation of galaxies at intermediate redshift ($z \sim$ 0.3), by mapping the detailed properties of stars and the ionised gas in galaxies. MAGPI will provide constraints on understanding the role of gas accretion and merging as well as tracing the metal mixing history of galaxies. It has a comparable \textit{physical} spatial resolution to local surveys \citep[$\sim$1$-$4~half-light radius/PSF FWHM;][]{Foster_2021} such as the Sydney-Australian-Astronomical-Observatory Multi-object Integral-Field Spectrograph \citep[SAMI;][]{Croom_2021} and Mapping Nearby Galaxies at Apache Point Observatory \citep[MaNGA;][]{Bundy_2015} surveys.

    MAGPI uses the wide-field mode, yielding a $\sim 1 \times 1$ arcmin$^2$ field-of-view (FOV) sampled by $0.2 \times 0.2$ arcsec$^2$ spatial pixels (henceforth spaxels). The wavelength covers 4700 $\si{\angstrom}$ to 9350 $\si{\angstrom}$ and the spectral sampling is 1.25 $\si{\angstrom}$/spaxel. The GLAO system ensures that all MAGPI targets are observed with an effective seeing of 0.65 arcsec FWHM in $V$-band, or better. For each primary target, the total on-source integration time per field is 4.4~hours, which ensures a signal-to-noise ratio of 5 $\si{\angstrom}^{-1}$ per resolution element around 6000\AA\ - 6500\AA\ in the stellar continuum for individual spaxels at the half-light radius (\re).

    The data reduction process is briefly described in \citet{Foster_2021} and will be provided in further detail in Mendel et al. (in prep). 
    The raw data are processed using {\sc{pymusepipe}} \footnote{\url{https://github.com/emsellem/pymusepipe}} \citep{Weilbacher_2020}. 
    The final processing of the individual MUSE science exposures is performed outside of the standard pipeline, using {\sc{cubefix}} \citep{Cantalupo_2019} and Zurich Atmosphere Purge sky-subtraction software \citep{Soto_2016}.
    The synthetic white-light (i.e., collapsed MUSE spectrum), $r$- and $i$-band images for each field are created using the {\sc{mpdaf}}\footnote{\url{https://github.com/musevlt/mpdaf}} python package.
    The {\sc{profound}} package \citep{Robotham_2018} is an image analysis package applied to the white-light images to detect objects and to produce a preliminary segmentation map. {\sc{profound}} is also used on $r$- and $i$-band images created from the MUSE datacubes for photometric parameters.

	\subsection{Sample Description}\label{subsec:sample}
\begin{figure*}
	    \centering
	    \includegraphics[width=0.495\textwidth, height=3.5in]{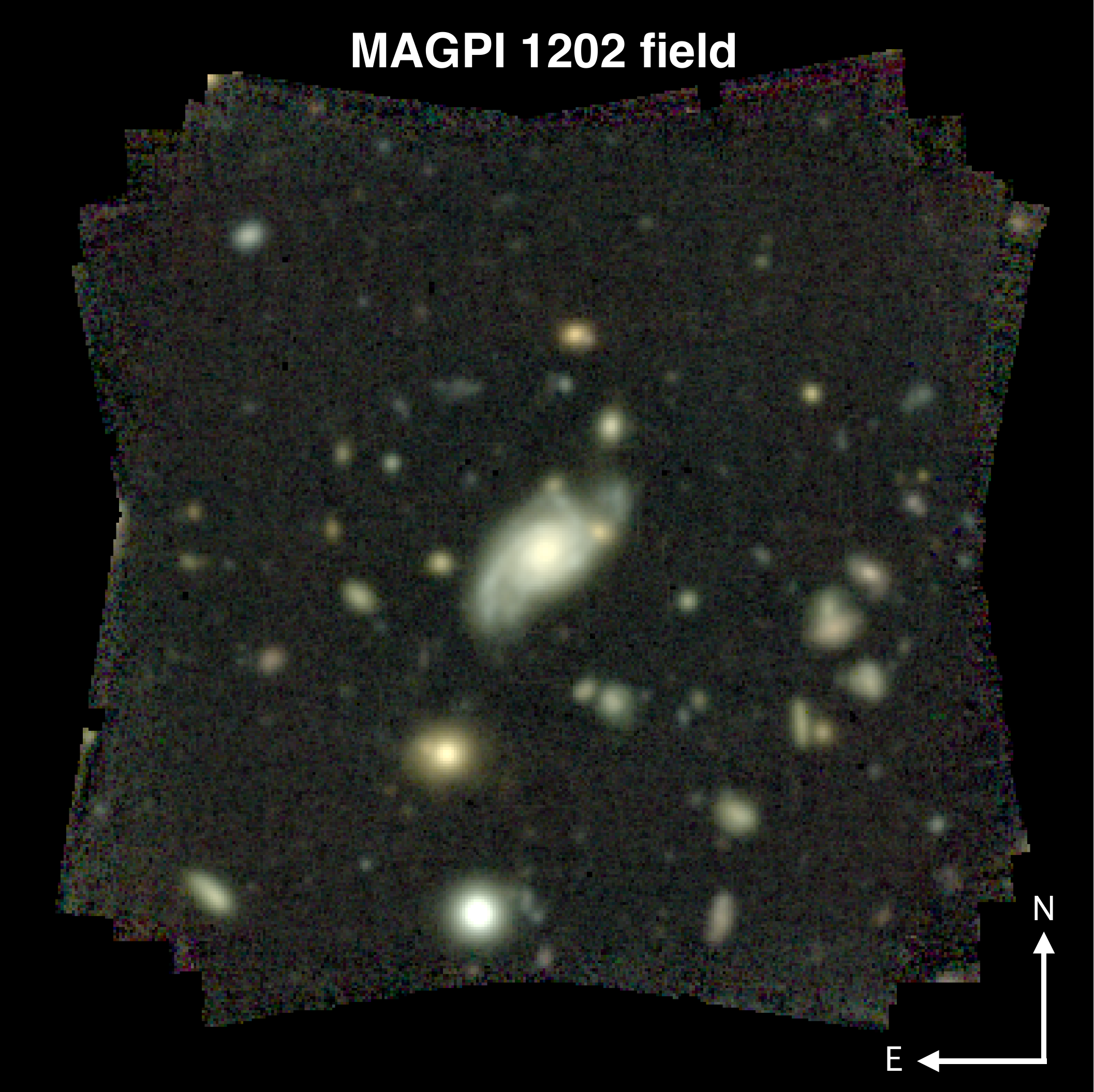}
	    \includegraphics[width=0.495\textwidth, height=3.5in]{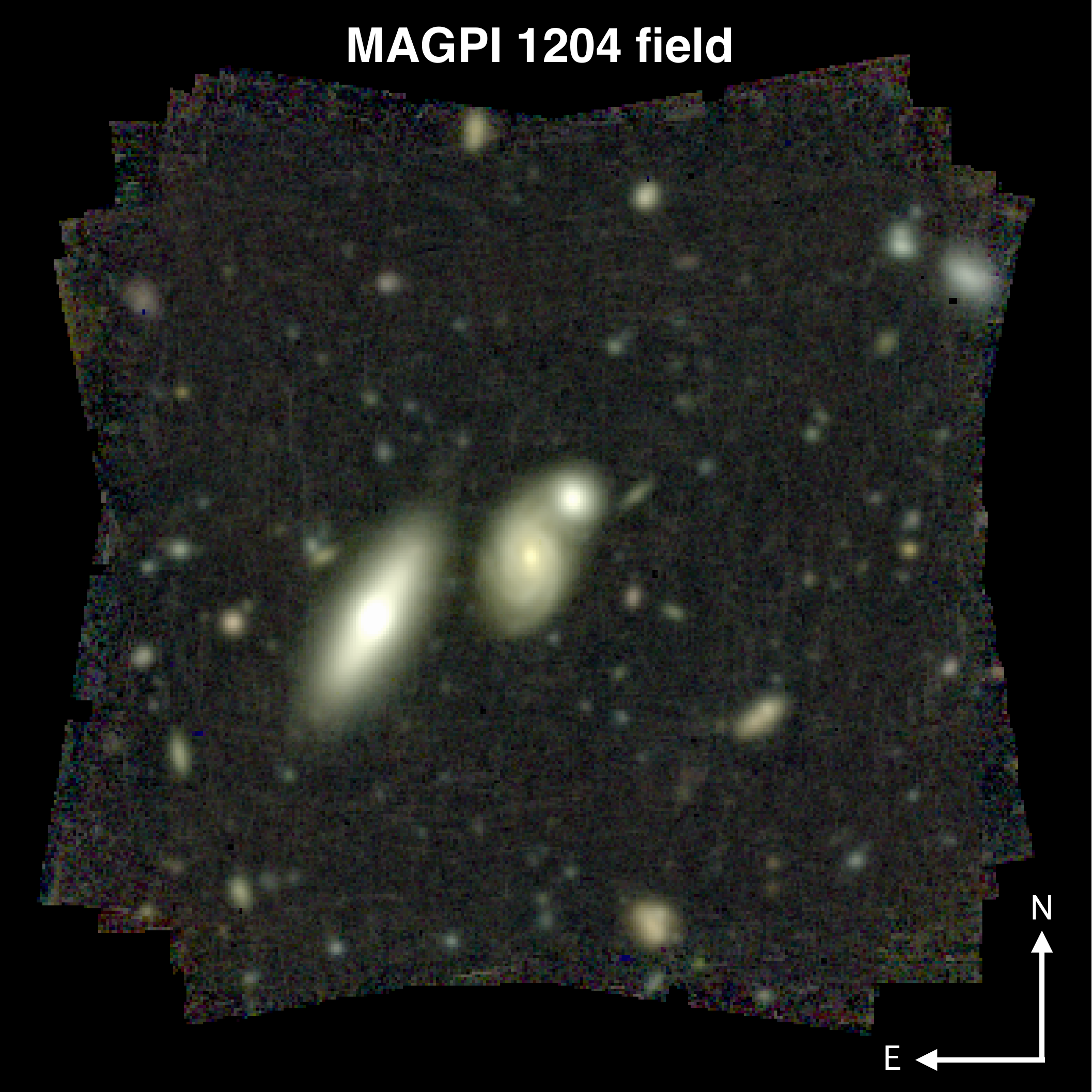}
	    \caption{MUSE $g_\mathrm{mod}ri$ \protect\footnotemark[4] composite images of the two well-defined spiral galaxies at the centre of each FOV of the MAGPI 1202 (left) and MAGPI 1204 (right) fields. Our pilot sample consists of MAGPI1202197197 (hereafter SG1202, at $z \simeq $ 0.2919) and MAGPI1204198199 (hereafter SG1204, at $z \simeq $ 0.3163), which are the large spiral galaxies located at the centre of each field. The companion galaxy to the northwest of SG1204 is a foreground galaxy at $z \simeq $ 0.254.}
	    \label{fig:FOV}
	\end{figure*}

        From the 35 observation-completed MAGPI fields (Oct 2022), we find 12 non-merging galaxies that prominently show grand-design or multi-armed spiral features from visual inspection. 
        We exclude merging spiral galaxies to remove complexities that could be introduced to the stellar and gas properties from tidal interactions.
        Tidally-driven spiral arms are not discussed in this work.
        Fig~\ref{fig:FOV} shows the coloured images of our pilot sample including two three-armed face-one spiral galaxies: MAGPI1202197197 (left of Fig~\ref{fig:FOV}, hereafter SG1202) and MAGPI1204198199 (right of Fig~\ref{fig:FOV}, hereafter SG1204).
        The pilot sample has the largest \re\ and the highest signal-to-noise ratio (SNR) among the 12 non-merging spiral galaxies, which allows for a detailed study and comparison between arm and interarm regions without the need for binning.
        This paper aims to demonstrate the feasibility of tracing spectral parameters and their linkage to spiral structures at $z \sim 0.3$. 
        In our next paper, we will apply the methodology to the full MAGPI survey upon the completion of observation.
        The two spiral galaxies are part of a group according to the Galaxy and Mass Assembly Galaxy Group Catalog \citep[G$^3$Cv1;][]{Robotham_2011} which indicates they are presumably under environmental effects from their companion galaxies.
        The physical parameters of galaxies in our pilot sample are listed in Table~\ref{tab:info}.
	
 \footnotetext[4]{$g_\mathrm{mod}$ covers part of the $g$ band redder than 4700\AA.}
	
	\begin{table*}
		\newcommand{\tabincell}[2]{\begin{tabular}{@{}#1@{}}#2\end{tabular}}
		\centering
            \small
		\begin{tabular}{ccccccccc}
			\hline
	    Galaxy    &  \tabincell{c}{R.A. \\ (degree)}  &  \tabincell{c}{Dec. \\ (degree)} &  \tabincell{c}{Inclination \\ (degree)}  &  \tabincell{c}{P.A. \\ (degree)}  &   Redshift   & log($M_*$/M$_{\odot}$)  &  \tabincell{c}{\re \\ (kpc)}  & pitch angle (degree)   \\
        (1) & (2) & (3) & (4) & (5) & (6) & (7) & (8) & (9)\\
	    \hline
	    MAGPI1202197197 (SG1202)   &  175.3388  & -1.5825 & 56.5 & 132.6 & 0.2920 & 10.88 $\pm$ 0.01 & 9.23 & 29.1$^{\circ}$, 27.5$^{\circ}$, 20.4$^{\circ}$\\
	    MAGPI1204198199 (SG1204)   &  175.6612  & -0.7943 & 46.9 & 175.0 & 0.3164 & 11.01 $\pm$ 0.01 & 10.09 & 16.7$^{\circ}$, 19.4$^{\circ}$, 21.0$^{\circ}$\\
	    \hline
	    \end{tabular}
	    \caption{Physical parameters of our pilot sample galaxies from MAGPI in this study. 
     {\bf{Column 1:}} Galaxy name in the MAGPI survey and their short name in this work. 
     {\bf{Column 2 \& 3:}} J2000 Coordinates of the galaxy in unit of degree. 
     {\bf{Column 4 \& 5:}} Inclination and photometric position angle (P.A.) are measured with {\sc{profound}} (Mendel in prep). 
     {\bf{Column 6:}} Redshift measured from {\sc{marz}}. 
     {\bf{Column 7:}} Stellar mass of the spiral galaxies from Mun et al. (submitted), applying single population synthesis to stellar energy distribution fitting on HSC images following \citet{Taylor_2011}. The total stellar mass is obtained by summing the stellar mass in each pixel within the dilated mask from {\sc{profound}}.
     {\bf{Column 8:}} Half-light radius (\re) determined with {\sc{profound}} on $i$-band images.
     {\bf{Column 9:}} Pitch angles of the spiral arms (see Sec~\ref{subsec:define_arm}) are listed clockwise from the north arm.}
	    \label{tab:info}
	\end{table*}

	\section{Data Analysis and Results}\label{sec:analysis}
        In this section, we present our analysis of the reduced MAGPI data including the spectral fitting processes, spaxel selection, the definition of spiral arms and inter-arm regions, and the measure of spectral parameters.
        Our goal is to constrain the origin of spiral features by studying the impacts of spiral arms on the gas ISM properties and stellar populations. 
        We focus on the two-dimensional distributions of star formation rate (Sec~\ref{subsec:SFR}), gas-phase metallicity (Sec~\ref{subsec:z}) and stellar age (Sec~\ref{subsec:age}) in the following analysis.
        
        \subsection{Emission line fits}\label{sec:gist}
        Two dimensional spectral maps are obtained with {\sc{gist}} \citep{Bittner_2019}, which is a pipeline with implemented {\sc{ppxf}} \citep[][]{Cappellari_2004, Cappellari_2017} and {\sc{pygandalf}} \citep{Sarzi_2006, Falcon-Barroso_2006} routines. 
        The {\sc{gist}} pipeline simultaneously extracts kinematics, emission lines and stellar population properties from full spectral fitting. 
        There are several unique customisations made (Mendel et al. in prep and Battisti et al. in prep) for its use with MAGPI data when fitting emission lines:

        \begin{itemize}
            \item A binning scheme is used where if the signal-to-noise ratio of the continuum is too low for the adopted threshold for Voronoi bins, then a single integrated bin is adopted instead. 
            Voronoi binning optimally preserves the maximum spatial resolution of two-dimensional data, given a constraint on the minimum signal-to-noise ratio \citep{Cappellari_2012}. The continuum fitting is based on binning scheme while the following emission line fits are on a spaxel level.
            \item A modified version of {\sc{pygandalf}} is implemented that:
                \begin{itemize}
                    \item employs three different sets of input parameters including the flux, velocity and velocity dispersion, which is adjusted based on the best fit of the stellar component. 
                    The fitting that yields the lowest $\chi ^2$ is considered as the final fitting results for the gas component.
                    This modification is tailored to effectively handle low S/N emission lines.
                    \item estimates the errors on fluxes, velocities, and standard deviations for the emission lines, based on a Monte Carlo approach.
                \end{itemize}
        \end{itemize}

        The emission line working group of the MAGPI survey adopts the light-weighted stellar templates from {\sc{ssp\_mist\_c3k\_salpeter}} (Charlie Conroy, private communication) and set the multiplicative Legendre polynomial to 12. 
        The emission lines are fit as a Gaussian function which is independent of the chosen IMF.
        The targeted SNR of the stellar continuum for Voronoi binning is 10. 
        The emission line fits are carried out with a single component tied to the brightest line for the wavelength range covered at the galaxy redshift (i.e., \ha\ for $0<z<0.424$, [O~\textsc{iii}]$\lambda 5007$ for $0.424<z<0.865$ and [O~\textsc{ii}]$\lambda 3729$ for $0.865<z<1.507$). The line velocity for each spaxel is restricted to $\pm$ 600 km/s relative to the stellar continuum velocity of the nearest Voronoi bin. 
        For each MAGPI field, Mendel et al. (in prep) empirically measure the line spread function (LSF), which will be adopted in the following fitting processes. 
        The spectral line data products, including [O~\textsc{ii}]$\lambda\lambda3726,29$, \dfth, H$\beta$ (4861\si{\angstrom}), [O~\textsc{iii}]$\lambda5007$, \ha\ (6563\si{\angstrom}) and [N~\textsc{ii}]$\lambda6584$ used in this work, are made by the MAGPI team and will be publicly available in a forthcoming data release (Battisti et al. in prep).

	\subsection{Spaxel selection criteria}\label{sec:snr_extinc_bpt}
\begin{figure*}
	    \centering
	    \includegraphics[height=2.8in]{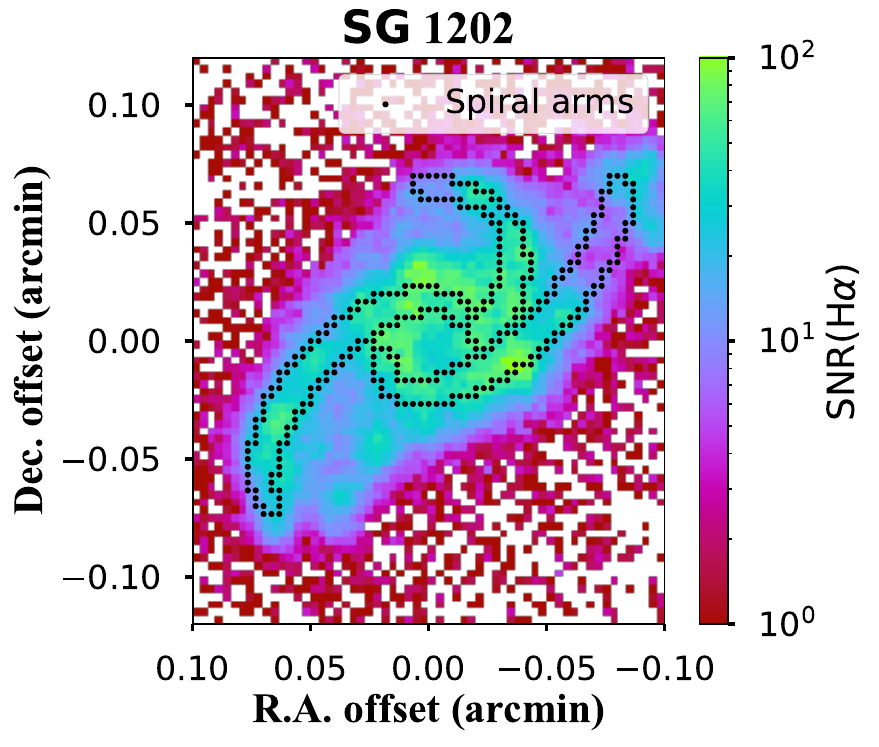}
	    \includegraphics[height=2.8in]{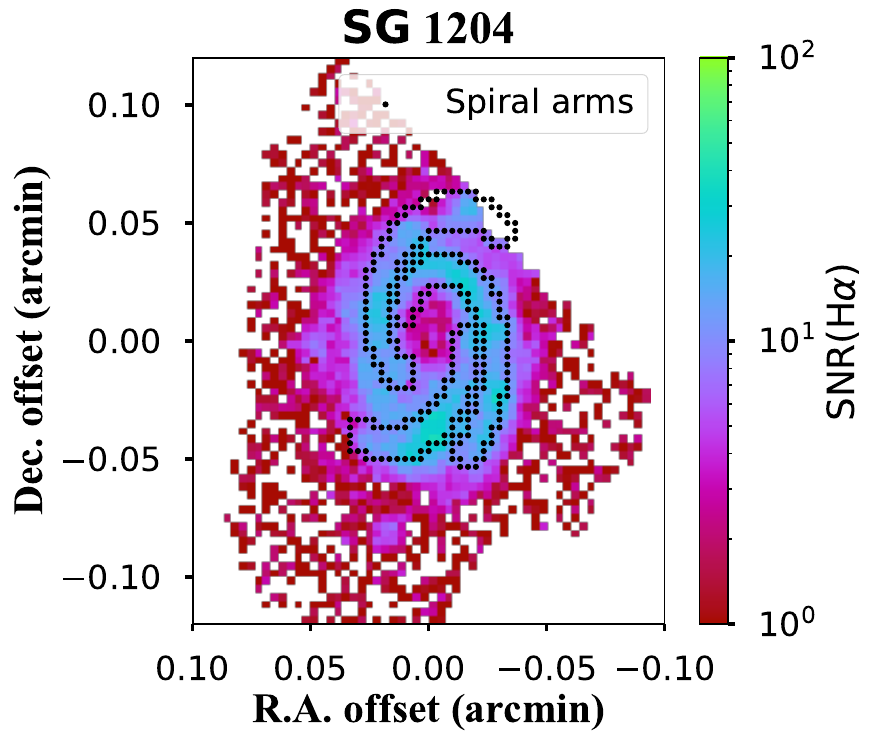}
	    \caption{SNR maps of SG1202 and SG1204. Only spaxels with SNR $\textgreater$ 3 in \ha\ and H$\beta$ are selected for all the following analyses. The defined spiral arms are over-plotted as black dots (see Sec~\ref{subsec:define_arm}). The spiral arms show a higher SNR of \ha\ than the inter-arm regions.
	    }
	    \label{fig:snr}
	\end{figure*}

	To achieve a reliable scientific analysis of SG1202 and SG1204, we apply SNR $\textgreater$ 3 limits (Fig~\ref{fig:snr}) to H$\beta$  and \ha\ emission lines before extinction correction. 
    The spaxels with SNR below 3 are excluded from all the following analyses. 
    To ensure sufficient spaxels and accurate analysis, we establish a 3$\sigma$ lower limit by assigning three times the noise as the new signal for [O~\textsc{iii}]$\lambda$5007 and [N~\textsc{ii}]$\lambda$6584 where if the spaxels have a SNR less than 3 \citep[see Sec~2.6 of][]{Rosario_2016}. 
    In face-on spiral galaxies, \ion{H}{ii} regions with low scale height destroy the dust within birth clouds and therefore are predominantly affected by dust in the foreground \citep{Wild_2011}. Thus, in this work, we apply the Milky Way extinction curve from \cite{Fitzpatrick_2019} as:
	\begin{equation}
	    E(B-V) = 2.5 \times \left( \frac{\log_{10}\frac{(H\alpha/H\beta)_{\mathrm{obs}}}{(H\alpha/H\beta)_{\mathrm{int}}}} {k_{H\beta}-k_{H\alpha}}\right).
	\end{equation}
	We adopt $R_{\nu}$ = 3.1 to determine the $k$ value at each wavelength and measure the colour excess $E(B-V)$ for all spaxels with SNR (\ha\ \& H$\beta$) \textgreater 3. The $E(B-V)$ values are then used to calculate the intrinsic emission line fluxes $F_{\mathrm{int}}$ from the observed fluxes $F_{\mathrm{obs}}$ as:
	\begin{equation}
	    F_{\mathrm{int}} = F_{\mathrm{obs}} \times 10^{0.4k_{\lambda} E(B-V)}.
	\end{equation}
	The extinction-corrected fluxes are used in all the following analyses to measure the gas-phase metallicities (Sec~\ref{subsec:z}).
	
	To limit our analyses to spaxels that arise predominantly from star formation (i.e., photoionisation) and exclude spaxels excited mainly by hard components (e.g., shocks, active galactic nuclei; AGN), we use the Baldwin, Phillips \& Terlevich diagram \citep[BPT diagram;][]{BPT_1981} to separate spaxels based on different excitation mechanisms. Based on $\mathrm{ [N~\textsc{ii}]\lambda6584/{H\alpha} }$ versus $\mathrm{ [O~\textsc{iii}]\lambda5007/H\beta }$, low-ionisation nuclear emission-line regions (LINERs) and AGN are distinguished from the \ion{H}{ii} regions, which are powered primarily by photoionisation \citep{Kewley_2001, Jin_2022}. 
	We separate AGN/shock spaxels from photoionisation by using the \cite{Kewley_2001} demarcation line, which models the starburst galaxies with PEGASE v2.0 and STARBURST99 to derive the theoretical classification scheme for AGN and \ion{H}{ii} regions. Another work \citep{Kauffmann_2003b} presents an empirical demarcation line based on the properties of $\sim$ 120,000 nearby galaxies with the Sloan Digital Sky Survey. 
	The spaxels beneath the \cite{Kauffmann_2003b} demarcation line are interpreted to be purely star-formation excited while data over the theoretical \cite{Kewley_2001} demarcation line predominately arises by AGN or LINER emission sources. A total of 1.1\% (0.8\%) spaxels in SG1202 (SG1204) lie above the \citet{Kewley_2001} line, which are excluded in the following analyses.
    	
	\subsection{Defining the Spiral Arms}\label{subsec:define_arm}
	We identify the spiral arms of our two galaxies with a combination of computer algorithms and manual checks, using the automated tool {\sc{sparcfire}} \citep{Davis_2014}, which was developed based on Galaxy Zoo data \citep{Lintott_2008, Lintott_2011}.
        To highlight the spiral features of our two galaxies at $z \sim$ 0.3, we employ {\sc{galfit}} to determine the optimal Sérsic component and subtract it from the white-light images. 
        Following the deprojection process in Sec~3.3 of \citet{Grasha_2017}, we deproject the galaxy images based on the inclination and P.A. in Tab~\ref{tab:info} which are then set as the input of {\sc{sparcfire}}. 
        The output of {\sc{sparcfire}} based on deprojected images is shown in Fig~\ref{fig:sparcfire}: red spiral arms are identified as having positive pitch angles (clockwise) while blue arms have negative pitch angles (counterclockwise).
        As three continuous spiral arms are vividly observable in both the white-light images (Fig~\ref{fig:FOV}), we only keep the three longest spiral arms from the output of {\sc{sparcfire}} for both spiral galaxies. 
	As a secondary step, we fine-tune the starting and ending radii of the spiral ridge lines to improve their alignment with the observed white-light images. 
	The final adopted spiral arm ridge lines are presented in Fig~\ref{fig:delta_phi}, overplotted on the \dphi\ map, which represents the angular azimuthal distance to the nearest spiral arm at a constant galactocentric distance (Sec~\ref{subsec:dphi}). 
    The typical spiral arm width in the Milky Way is $\sim$ 1~kpc \citep{Reid_2014}.
    We demarcate the width of spiral arms as 2~kpc in the MAGPI survey, taking the beam-smearing effects (0.6 $-$ 0.8 arcsec; FWHM) into account.
    We show the boundary of the spiral arms as black dots in Fig~\ref{fig:snr}.
	
    \subsection{Definition of azimuthal distance \dphi}\label{subsec:dphi}
    Depending on the physical mechanisms that assemble spiral arms, density wave theory or dynamic spirals, stellar populations and gas in the ISM between both sides of the spiral arms are expected to show different behaviours, which can be used to constrain the origin of the spiral features.
    The spatially resolved IFS data from our MAGPI observations allow us to compare the ISM properties between the spiral arm and inter-arm regions as well as differences between downstream and upstream in the inter-arm regions. 
    To identify the different parts of inter-arms, we define \dphi\ which measures the azimuthal distance to the nearest spiral arm. 
    We assign 
\begin{equation}
   \Delta \phi_i = -min(|\phi_j - \phi_i|)
\end{equation}
when pixel $i$ is on the leading edge of the nearest spiral arm and assign 
\begin{equation}
   \Delta \phi_i = min(|\phi_j - \phi_i|)
\end{equation}
when pixel $i$ is on the trailing edge of the nearest spiral arm. The pixel $i$ is the targeted pixel and the pixel $j$ is the pixel within the spiral regions at the same galactocentric distance (within 0.5~kpc uncertainty).
    
\begin{figure*}
	    \centering
	    \includegraphics[width=0.495\textwidth, height=3in]{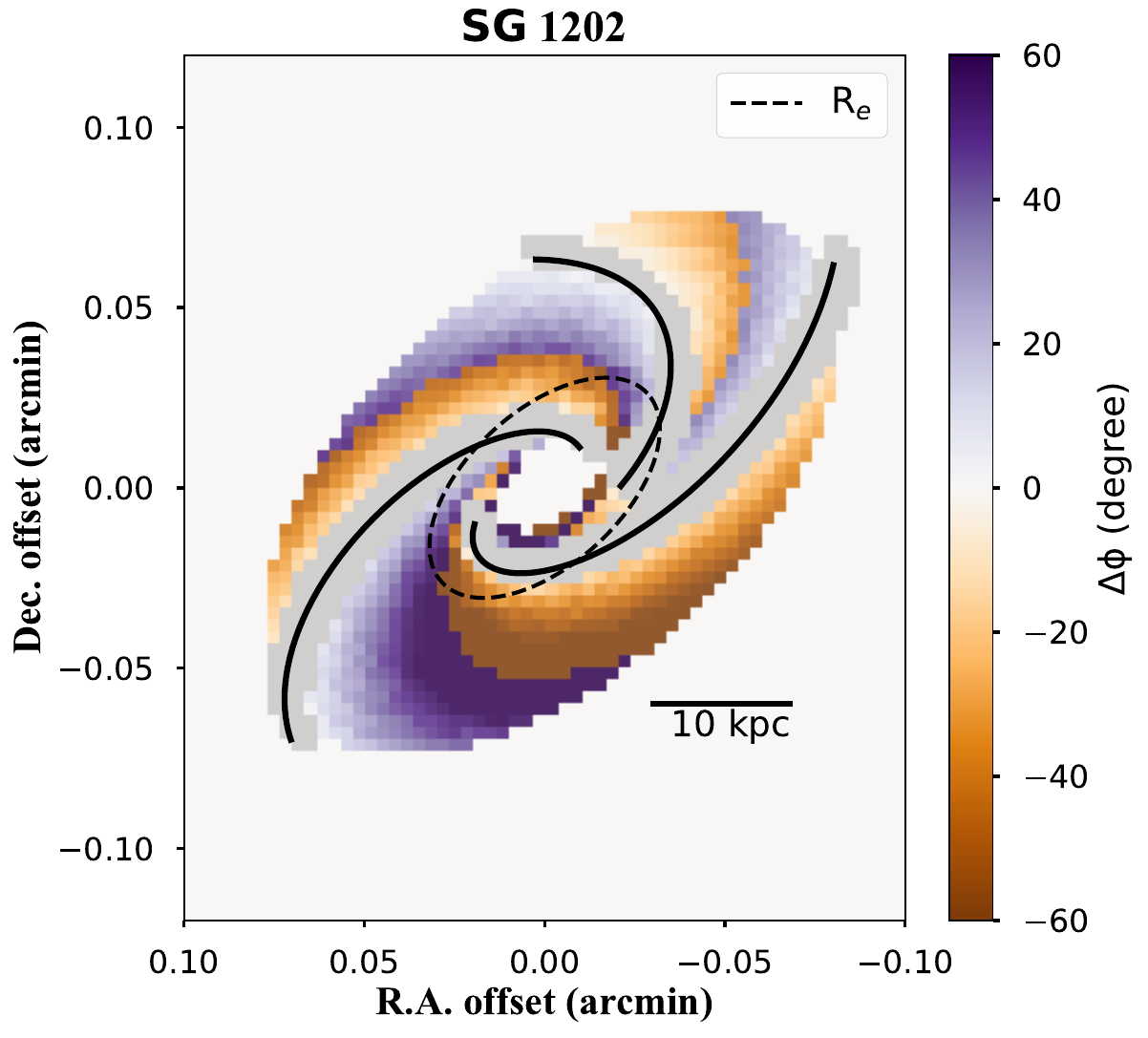}
	    \includegraphics[width=0.495\textwidth, height=3in]{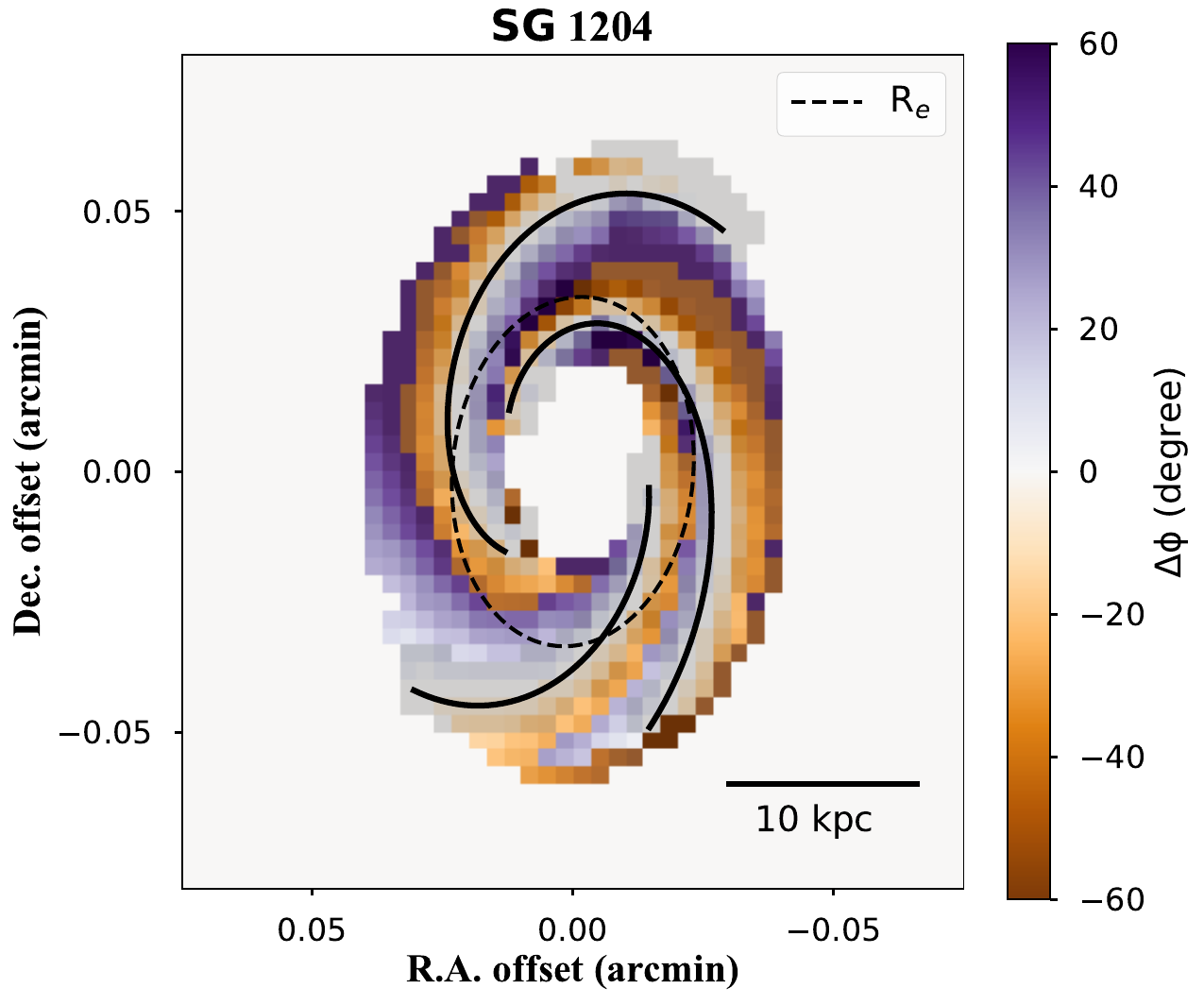}
	    \caption{A measure of the angular azimuthal distance to the nearest spiral arm (grey pixels) at a constant galactocentric distance (\dphi) of SG1202 (left) and SG1204 (right), over-plotted with the spiral arm ridge lines. 
        We define \dphi\ only for the spaxels located at the galactocentric distance where three spiral arms are present, thus there is an absence of \dphi\ in the central region.
        Inter-arm spaxels with negative (positive) \dphi\ are classified as downstream (upstream), colour-coded as orange (purple). 
        1\re is shown as the dashed ellipse.
	    }
	    \label{fig:delta_phi}
	\end{figure*}

    With the defined spiral arm ridge lines, the \dphi\ maps of SG1202 and SG1204 are presented in Fig~\ref{fig:delta_phi}.
    The inter-arm spaxels with negative (positive) \dphi\ are classified as downstream (upstream). 
    In the following sections, we will analyse ISM properties and stellar populations as tests of the density wave theory.
    
	\subsection{Star formation rate}\label{subsec:SFR}
	Star formation is one of the major drivers of galaxy evolution and is the process that enriches the ISM with metals. 
    Young, massive stars produce copious amounts of ionising photons that are observed in emission nebular lines (e.g., \ha). Stars more massive than $\sim$ 10~$\mathrm{M_\odot}$ produce a measurable ionising photon flux and only live $\sim$ 10~Myr, providing a sensitive indicator of young, massive stellar populations. 
        In this work, we adopt the star formation rate (SFR) prescription from \citet{Kennicutt_1998} to measure the SFR for each spaxel:
	\begin{equation}
	    \mathrm{ SFR (M_\odot yr^{-1}\ cm^{-2}) = 7.9 \times 10^{-42} \times F_{H\alpha} (erg\ s^{-1}\ cm^{-2}) },
	\end{equation}
	where $F_{\mathrm{H\alpha}}$ is the flux of \ha\ per spaxel after dust-correction (Sec~\ref{sec:snr_extinc_bpt}).
        For consistency with the MAGPI survey, we adopt the \citet{Chabrier_2003} IMF.
        Following \cite{Bernardi_2010}, the SFR is divided by 10$\mathrm{^{0.25}}$ to convert the IMF from the original \citet{Salpeter_1955} to our preferred \citet{Chabrier_2003} IMF.
	
\begin{figure*}
	    \centering
	    \includegraphics[height=2.5in]{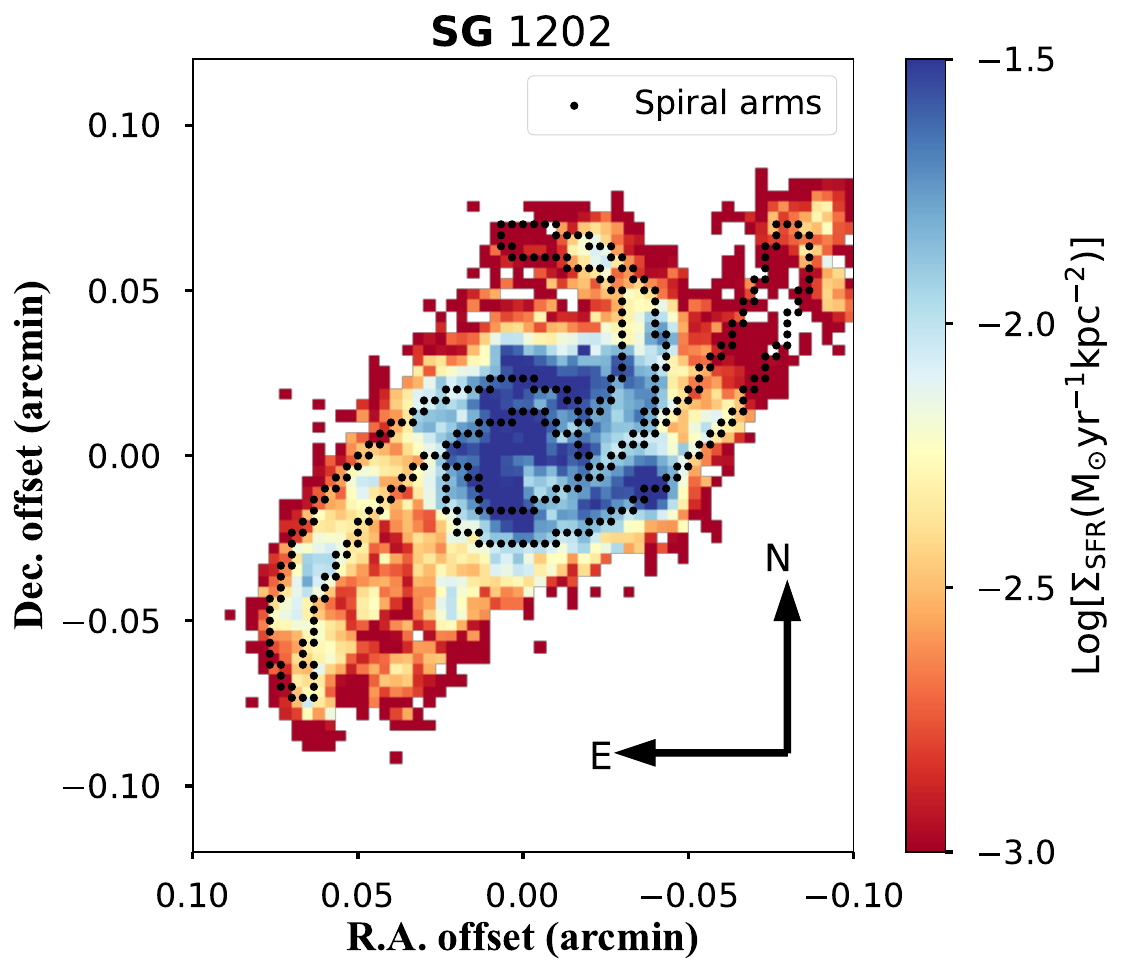}
	    \includegraphics[height=2.5in]{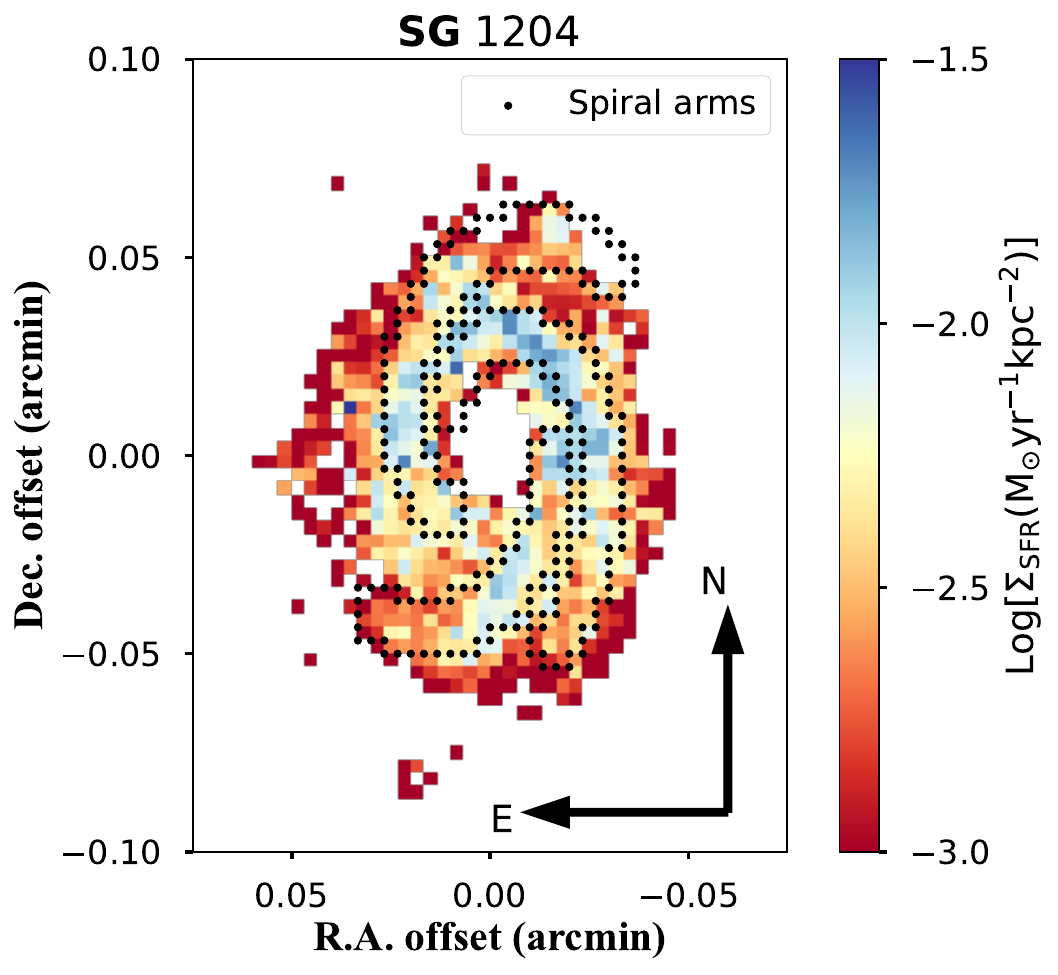}
	    \caption{\SigSFR\ maps based on dust-corrected \ha\ flux of SG1202 (left) and SG1204 (right), over-plotted with the boundary of defined spiral arm regions. Higher (blue) \SigSFR\ is predominantly located within the arm regions.
	    }
	    \label{fig:sfr}
	\end{figure*}

	We calculate the spatially resolved SFR surface density (\SigSFR) as follows:
        \begin{equation}
            \mathrm{ \Sigma_{SFR} (M_\odot yr^{-1} kpc^{-2}) = \frac{SFR \times \ 10^{-0.25}}{[D_A (kpc) \times 0.2] ^2} },
        \end{equation}
        where $D_\mathrm{A}$ is the angular diameter distance of the galaxy and 0.2 arcsec pixel$^{-1}$ is the pixel scale of the MAGPI survey.  
        The higher \SigSFR\ in the spiral arms than in the inter-arm regions observed in both SG1202 and SG1204 shows that the enhanced star formation is occurring within the spiral arms (see derived \SigSFR\ maps in Fig~\ref{fig:sfr}).

\begin{figure*}
	    \centering
	    \includegraphics[height=2.6in]{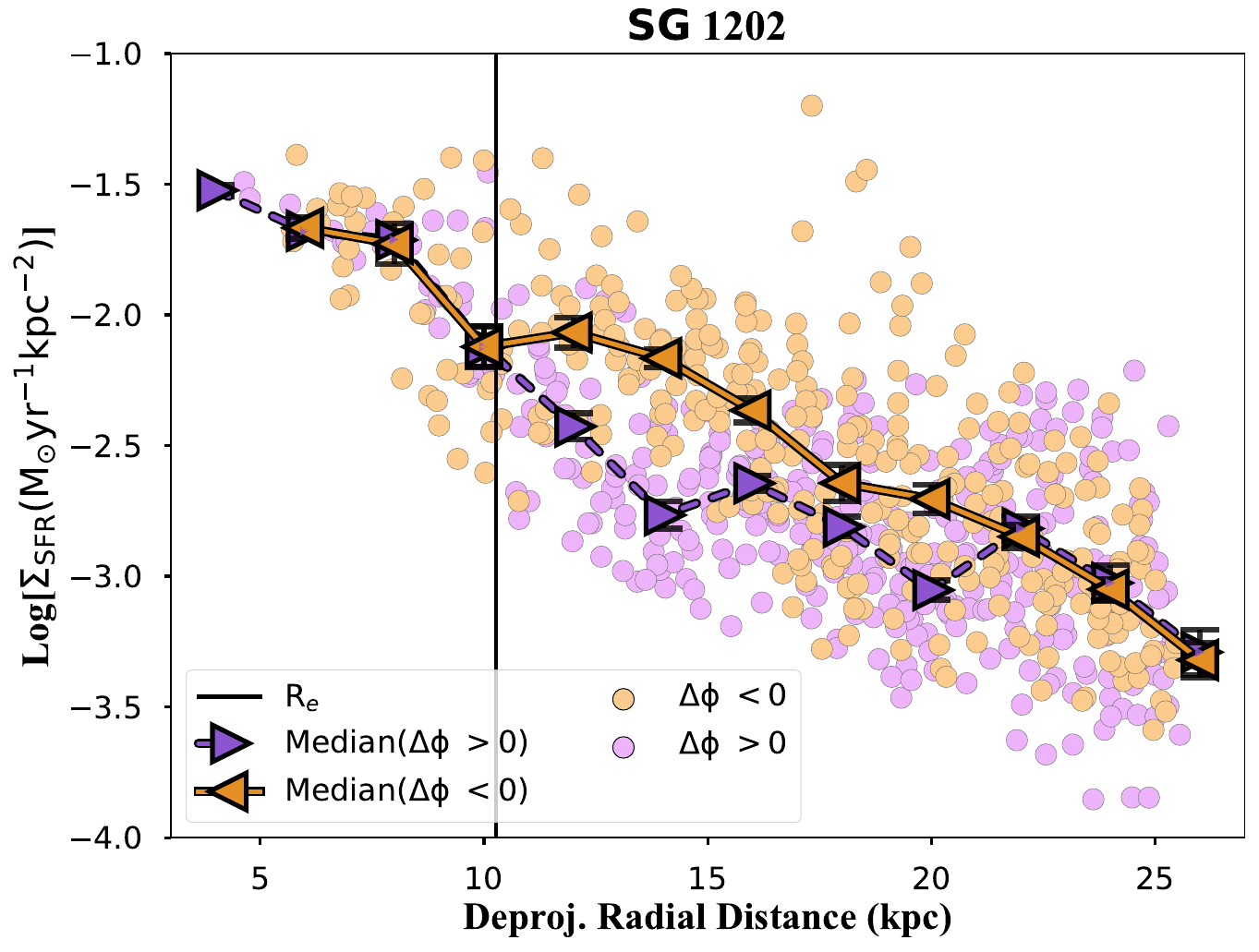}
	    \includegraphics[height=2.6in]{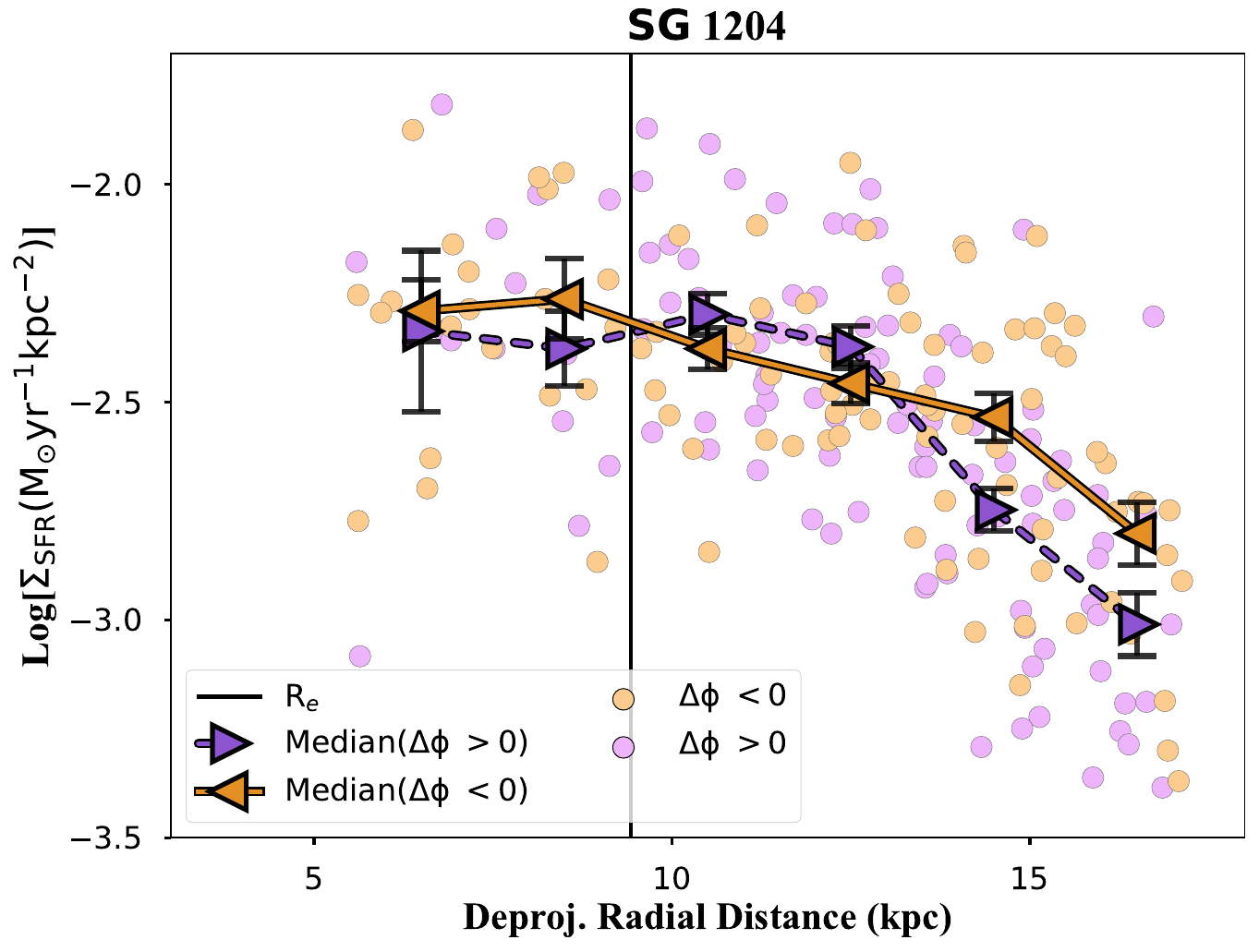}
	    \caption{Radial profiles of \SigSFR\ of SG1202 (left) and SG1204 (right), colour-coded by their location in the spiral galaxy: \dphi\ $< 0$ (downstream; orange) and \dphi\ $> 0$ (upstream; purple). 
     The $x$-axis is the radial distance to the galaxy centre after deprojection \citep[following][]{Grasha_2017}.
     The medians of each 1~kpc elliptical bin are marked as triangles with 1~$\sigma$ error bars. The vertical solid line marks the location of 1\re\ in each galaxy.
     Significantly higher \SigSFR\ is observed in the downstream (\dphi\ $< 0$; orange) of SG1202 than the upstream (\dphi\ $> 0$; purple) between 1 $\textless$ R/\re $\textless$ 2, whereas SG1204 shows comparable \SigSFR\ in the \dphi\ $< 0$ (downstream) and \dphi\ $> 0$ (upstream) regions across all galactocentric distances. 
	    }
	    \label{fig:sfr_updown}
	\end{figure*}

	Fig~\ref{fig:sfr_updown} compares the radial profiles of \SigSFR\ in the \dphi\ $< 0$ (downstream; orange) and \dphi\ $> 0$ (upstream; purple) regions as a function of deprojected radial distance. 
    In SG1202 (left panel of Fig~\ref{fig:sfr_updown}), we observe a generally higher \SigSFR\ in the downstream (orange) than in the upstream (purple) between 1 and 2\re\ ($\sim$ 10~kpc $-$ 20~kpc). 
    The \SigSFR\ in the downstream (\dphi\ $< 0$) of SG1204 (right panel of Fig~\ref{fig:sfr_updown}) is indistinguishable from the upstream (\dphi\ $> 0$) at all galactocentric distances and has no noticeable change at 1\re, unlike SG1202.
	The \SigSFR\ radial trends in different \dphi\ of SG1202 are consistent with the theoretical expectations from the quasi-stationary density wave theory (Sec~\ref{sec:intro}).
    Our data indicate more intense star formation located on the leading side of the spiral arm, compared to the trailing side between 1 and 2\re.
    The absence of an offset in \SigSFR\ across all radii of SG1204 can be the consequence of the galaxy having a tighter pitch angle (last column in Tab~\ref{tab:info}) compared to SG1202. 
        This scenario is further discussed in Sec~\ref{sec:discussion} and requires a larger observational sample to investigate.

	\subsection{Gas-phase metallicity}\label{subsec:z}
    The spatial distribution of gas-phase metallicities is a critical component for understanding the physical evolution of spiral galaxies, including their star formation history and mixing processes in the ISM \citep{Maiolino_2019, Kewley_2019, Li_2021, Sharda_2023}.
    Oxygen is the most abundant metal element in the Universe, which is mostly produced on short timescales by Type II supernovae along with other $\alpha$-elements. 
    In this work, we measure the oxygen abundances to represent the gas-phase metallicities of our galaxies utilising the N2O2 diagnostic \citep[][]{KD_2002}. 
    The N2O2 diagnostic uses three bright optical lines: [N~\textsc{ii}]$\lambda$6584/[O~\textsc{ii}]$\lambda\lambda$3726,3729. 
    The combination of [N~\textsc{ii}]$\lambda$6584 and [O~\textsc{ii}]$\lambda\lambda$3726,3729 requires an reliable extinction correction which is carried out in Sec~\ref{sec:snr_extinc_bpt}.
    The N2O2 diagnostic is insensitive to the ionisation parameter or ionising spectrum hardness \citep{Zhang_2017, Kewley_2019}. As such, it is also insensitive to the diffuse ionised gas that permeates outside of \ion{H}{ii} regions. 
    It makes N2O2 an ideal metallicity diagnostic for our observations. 
    We utilise the python package {\sc{pymcz}} \citep[][]{BIANCO201654} to measure the gas-phase metallicities and uncertainties with a Monte Carlo method. 

\begin{figure*}
	    \centering
	    \includegraphics[width=0.99\textwidth]{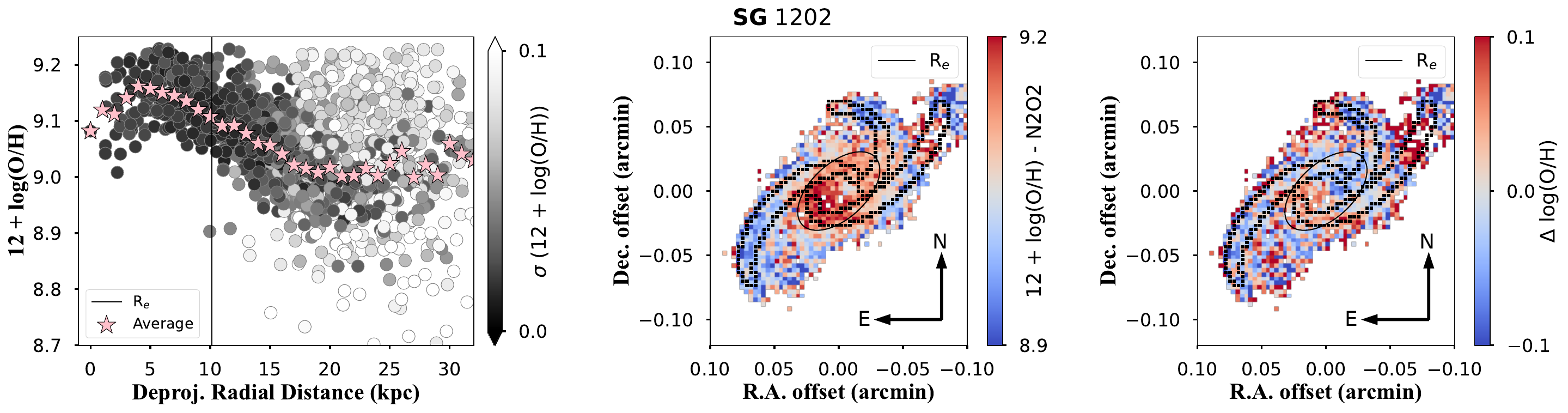}
	    \includegraphics[width=0.99\textwidth]{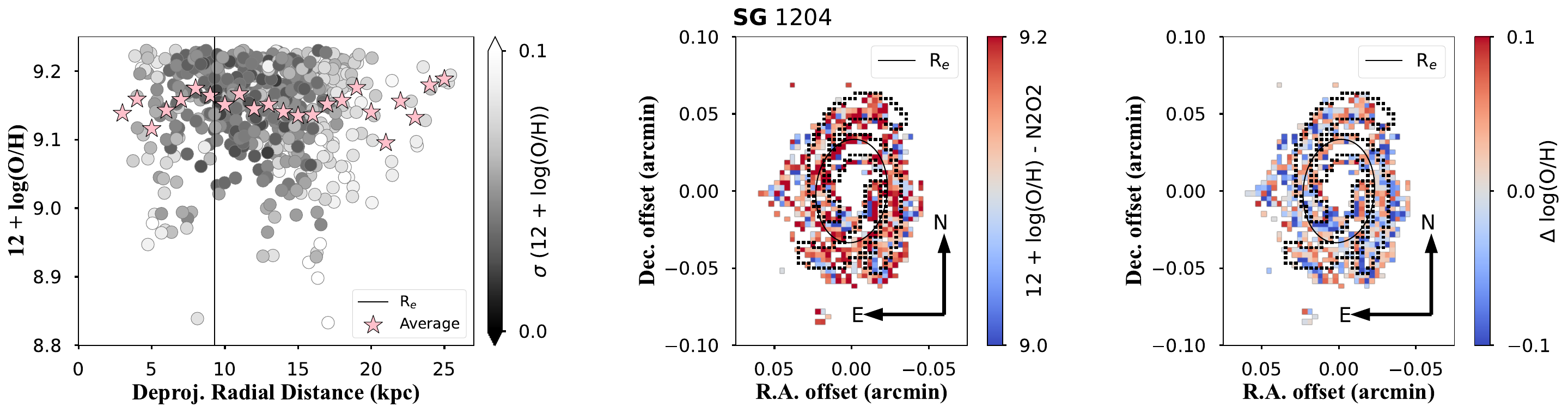}
	    \caption{\textbf{Left panels:} Radial profiles of gas-phase metallicity of SG1202 (top) and SG1204 (bottom), color-coded by 1$\sigma$ uncertainty. The vertical solid line delimits the 1\re. The pink star symbols are the weighted average of each 1~kpc elliptical bin. 
        We observe a steep-shallow metallicity gradient in SG1202 which is consistent with the simulation in \citet{Garcia_2022} and local observations in \citet{Grasha_2022, Chen_2023}. We find a flat metallicity gradient in SG1204 which indicates strong mixing effects. \textbf{Middle panels:} gas-phase metallicity maps 12 + log(O/H) of SG1202 (top) and SG1204 (bottom), overplotted with the boundary of spiral arms and an ellipse at the location of 1\re. We find a concentration of blue spaxels with low metallicity within the spiral arms of SG1202 but no systematic difference between arms and inter-arms of SG1204. \textbf{Right panels:} residual metallicity maps \offset\ of SG1202 (top) and SG1204 (bottom) obtained by subtracting the radial gradient (pink star symbols in the left column) from the metallicity maps (middle), overplotted with the boundary of spiral arms and ellipse at 1\re. We observe a lower \offset\ in the arm regions of SG1202, but no significant azimuthal variation in SG1204. This difference can be a consequence of efficient mixing which is implied in the flat metallicity gradient of SG1204.
	    }
	    \label{fig:z}
	\end{figure*}

    We present the metallicity radial profiles, colour-coded by their 1$\sigma$ uncertainty, in the left column of Fig~\ref{fig:z}. We show the 2D metallicity maps in the middle column of Fig~\ref{fig:z}.
    The observed negative radial gradient ($-0.069\pm0.002$ dex/$R_e$) in SG1202 is indicative of typical inside-out formation, commonly seen in local spiral galaxies \citep{Sanchez-Menguiano_2018, Chen_2023} while the flat metallicity gradient ($-0.014\pm0.007$ dex/$R_e$) in SG1204 suggests strong gas mixing which could be induced by recent interactions or cosmic gas accretion \citep{Kewley_2010, Rupke_2010a, Rupke_2010b, Perez_2011, Vollmer_2012}. 
    The shallower gradient in the \SigSFR\ profile of SG1204 compared to the one of SG1202 is also suggestive of a stronger mixing process, which can cause the absent azimuthal variation in the metallicity of SG1204 (further discussed in Sec~\ref{sec:discussion}).

    From the 2D metallicity map of SG1202 (top middle panel of Fig~\ref{fig:z}), we observe significantly higher metallicities (redder colour) in the inner region which is consistent with the negative metallicity gradient. We notice bluer spaxels (low metallicities) concentrated within the spiral arms than the inter-arms.
    In the 2D metallicities map of SG1204 (bottom middle panel of Fig~\ref{fig:z}), we find comparable metallicity across the disc region which is in line with the observed shallow metallicity gradient. We do not find a distinctive difference in the metallicity between spiral arms and inter-arms which could be attributed to the tighter spiral arms.

    Azimuthal metallicity variation can inform us of how metals are mixed in with their neighbouring ISM when gas and stars are rotating in their orbits \citep{Ho_2017}. 
    The IFS data of MAGPI allow us to study small-scale variations (0.6 $-$ 0.8 arcsec FWHM) present in the gas-phase metallicity, which is typically a much weaker trend than the global radial gradient.
    To calculate the residual metallicity maps \offset, we first calculate the weighted averages of each 1~kpc elliptical bin to represent the radial gradients and subtract that from the metallicity maps.
    Positive (negative) residual \offset\ indicates higher (lower) 12 + log(O/H) compared to spaxels at the same galactocentric distance; this allows us to study the azimuthal variations, if present.
	
	In SG1202 (top right panel of Fig~\ref{fig:z}), we find higher \offset\ in the inner region of the northwest spiral arm while the north arm and east arm are systematically dominated by negative \offset. 
    It is challenging to directly compare the metallicity between the arm and inter-arm regions based on the \offset\ map of SG1204 (lower panel of Fig~\ref{fig:z}), since the scale of \offset\ is comparable to the typical uncertainty of gas-phase metallicity (0.1~dex).
    We perform a Kolmogorov–Smirnov (KS) test to compare the distributions of \offset\ in the arm and inter-arms. The results of the KS test and the $p$-values are shown in Table~\ref{tab:ks} and summarised below. 
    The $p$-value measures the probability of the null hypothesis that the two samples are drawn from the same distribution for which a significance level of 1\% is commonly used. 
    It is worth mentioning that the $p$-values are an over-representation of how different the distributions truly are because the large sample size will cause the KS test to have very high power for discerning small differences in the distributions. 
    In addition, the KS test does not take into account the error in the parameters \citep{Lazariv_2018}.
 
    For SG1202, the $p$-value (1.7 $\times 10^{-4}$) of the KS test suggests that \offset\ has a different distribution in the arms compared to the inter-arms. 
    We find generally negative \offset\ in the spiral arms with an error-weighted average of -3.9$\pm 1.0 \times 10^{-3}$. 
    This is lower than the inter-arms with an error-weighted average of 4.3$\pm 1.0 \times 10^{-3}$.
    SG1204 has spiral arms that are significantly tighter (with a smaller pitch angle) than SG1202. The KS test suggests that it is possible that the \offset\ distributions in the arms and inter-arms are drawn from the same distribution (with a $p$-value of 6.8 $\times\ 10^{-2}$; Tab~\ref{tab:ks}). 
    The difference in the \offset\ of SG1202 and SG1204 is implied in the more wounded spiral arms of SG1204, coming along with stronger mixing processes. We discuss the physical mechanisms driving this difference in detail in Sec~\ref{sec:discussion}.
	
\begin{figure*}
	    \centering
	    \includegraphics[height=2.6in]{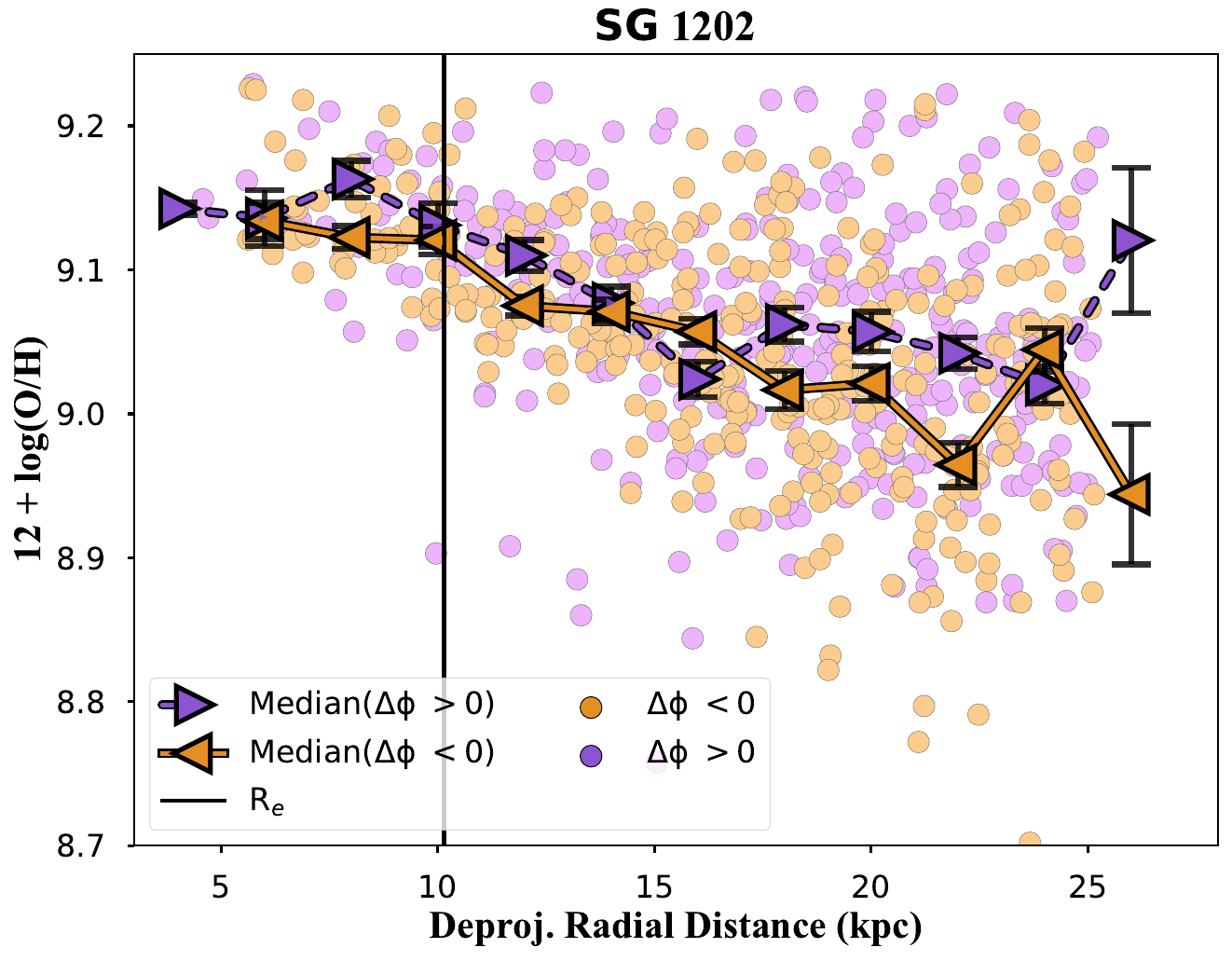}
            \includegraphics[height=2.6in]{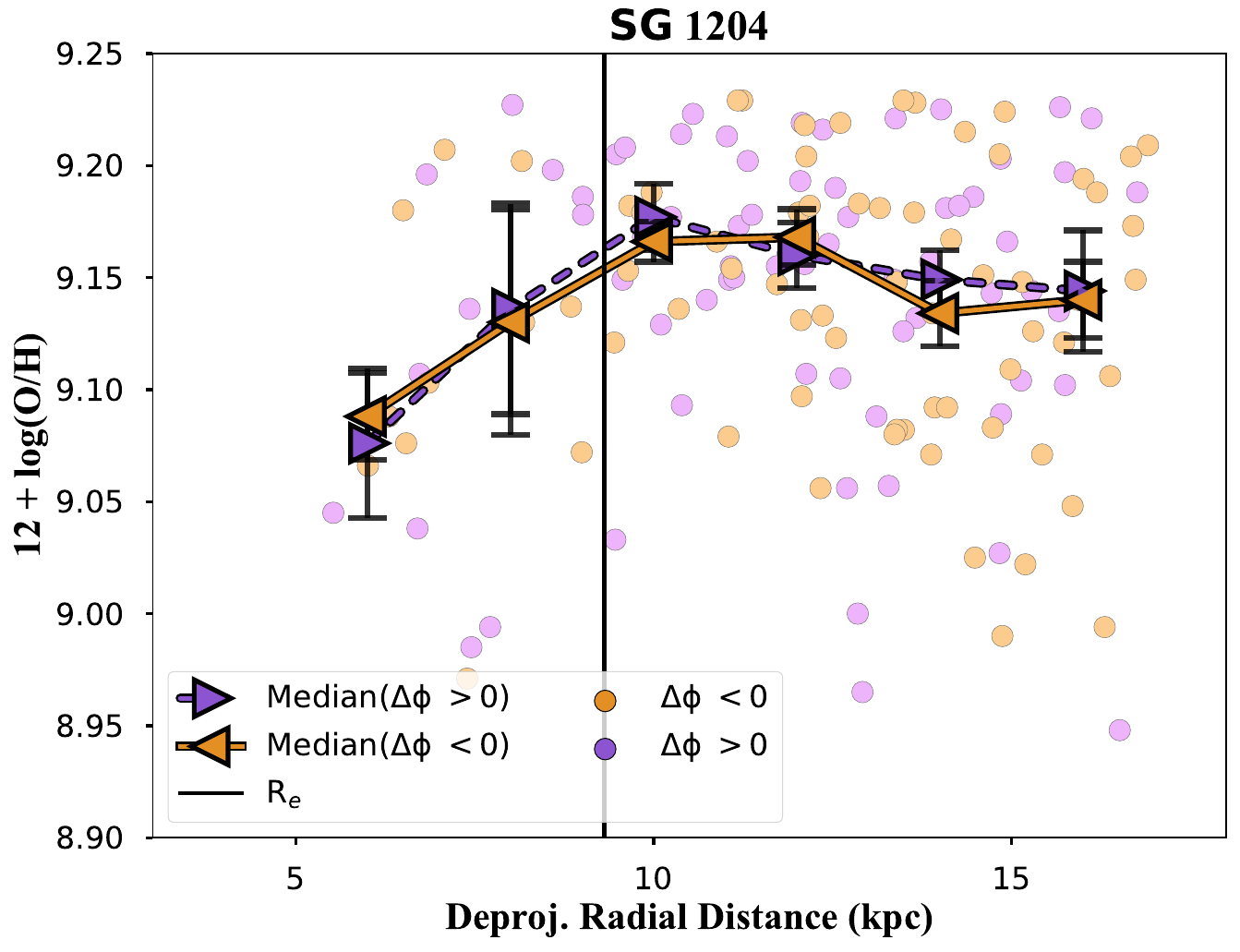}
	    \caption{Comparison of metallicity in the \dphi\ $< 0$ (downstream; orange) versus \dphi\ $> 0$ (upstream; purple) spaxels in SG1202 (left) and SG1204 (right). The large solid triangles are the medians of each 1~kpc elliptical bin, with 1$\sigma$ error bars. 1\re\ is marked as vertical solid lines. 
        The KS test suggests the 12 + log(O/H) in the downstream and upstream of SG1202 are possibly drawn from different distributions while agreeing that the 12 + log(O/H) on both sides of the spiral arms in SG1204 are possibly drawn from the same distribution.
     }
	    \label{fig:z_updown}
	\end{figure*}

        Based on the measurement of \dphi\ (Sec~\ref{subsec:dphi}), we compare the 12 + log(O/H) on different sides of the spiral arms across different radii (Fig~\ref{fig:z_updown}). 
        In SG1202, we observe marginally higher metallicities in the upstream (purple; \dphi\ $> 0$) than in the downstream (orange; \dphi\ $< 0$), outside 1$\sigma$ uncertainty and within 3$\sigma$ uncertainty. 
        The $p$-value (4.7 $\times$ 10$^{-3}$) from the KS test strongly rejects the null hypothesis that the 12 + log(O/H) in the downstream and upstream are drawn from the same distribution. 
        The observed azimuthal variation in the metallicity of SG1202 suggests that the origin and development of spiral arms in this galaxy are consistent with the density wave theory.
        However, we find that the metals in the ISM are well mixed in the inter-arms of SG1204. The $p$-value ($4.1 \times 10^{-1}$) from the KS test supports the same distribution of 12 + log(O/H) on both sides of the spiral arms of SG1204.
        The absence of azimuthal metallicity variation in SG1204 is consistent with the dynamically-driven spiral arms.
        Besides, the different azimuthal metallicity distributions in the SG1202 and SG1204 can be attributed to the tighter pitch angle of SG1204 which implies a shorter (more efficient) mixing time scale and higher ISM mixing effects.

	\subsection{Stellar age}\label{subsec:age}
        In addition to the distribution of the ISM gas, the distributions of stellar populations also hint at the origin and evolution of spiral galaxies.
	In this work, we constrain the stellar age with \dfth, which is a continuum feature that occurs around 4000\AA\ due to the absorption in the atmosphere within older stellar populations \citep{Noll_2009}.
    Due to our limited spectral quality, we are unable to operate spectrum energy distribution (SED) fitting on a spaxel scale.
    To get a sufficient SNR for SED fitting, we must bin the MAGPI data into a coarser spatial resolution, rendering the spiral arms and inter-arm regions no longer distinguishable.
	Based on the {\sc{gist}} stellar continuum fits (Sec~\ref{sec:gist}), we determine the \dfth\ value on a spaxel level following:
        \begin{equation}
            \mathrm{D_{4000} = \frac{Flux_{restframe}(4050\mathring{A} - 4250 \mathring{A} )}{Flux_{restframe}(3750\mathring{A} - 3950 \mathring{A} )}}.
            \label{equ:d4k}
        \end{equation}

    We exclude spaxels with \dfth\ uncertainty larger than 0.5 in the following analysis of stellar age.

\begin{figure*}
	    \centering
	    \includegraphics[width=0.9\textwidth]{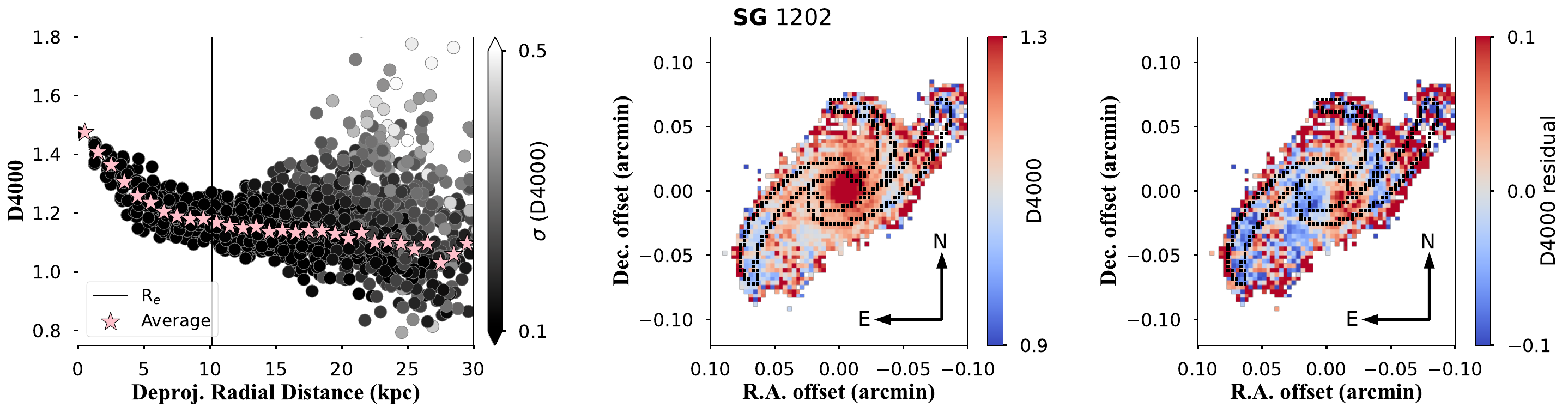}
	    \includegraphics[width=0.9\textwidth]{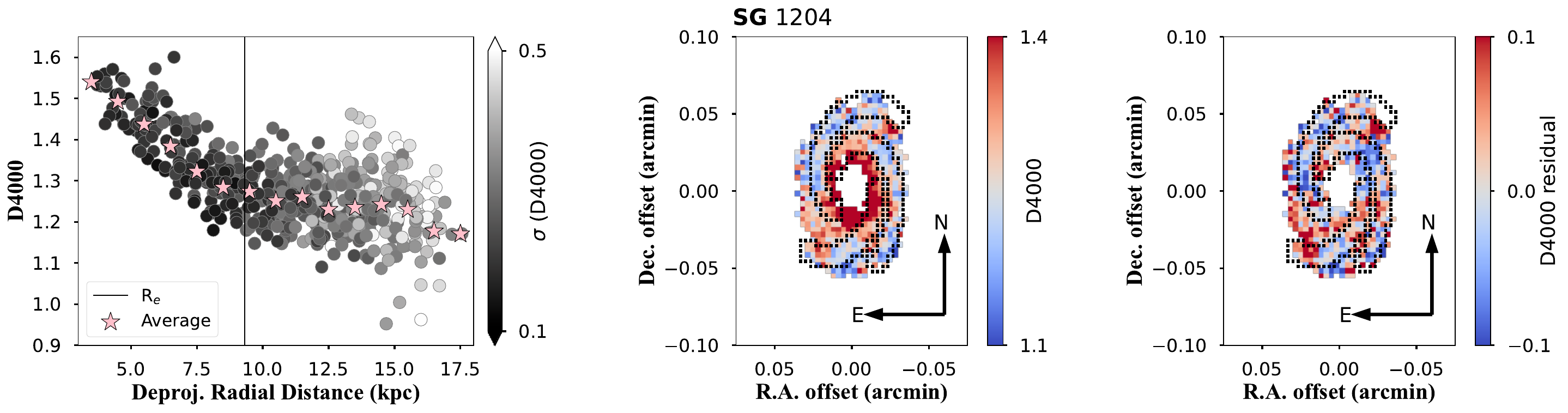}
	    \caption{Radial profiles colour-coded by 1$\sigma$ uncertainty (left), 2D maps (middle), and residual maps (right) of \dfth\ in SG1202 (top) and SG1204 (bottom). The residual maps are calculated by subtracting the weighted average of each 1~kpc bin (light pink star symbols in the left panel) from the observed \dfth\ maps (middle panel). The location of \re\ is shown as a vertical line in the left panel and as an ellipse in the middle and right panels. The boundary of the spiral arm regions is over-plotted as dashed lines on the \dfth\ and residual maps. We find negative \dfth\ radial gradients in both galaxies which support an inside-out formation scenario. Both \dfth\ maps and residual maps indicate lower \dfth\ (younger stellar populations) in the spiral arms of SG1202 and SG1204.
	    }
	    \label{fig:d4k}
	\end{figure*}

	Consistent with inside-out galaxy formation, we find negative gradients in the \dfth\ of SG1202 and SG1204 (left panels of Fig~\ref{fig:d4k}), with a flattening trend outside 1\re. In the middle panels of Fig~\ref{fig:d4k}, we present the 2D maps of \dfth, overplotted with the boundary of the spiral arms.
    In SG1202, we observe slightly lower \dfth\ (bluer spaxels) in the spiral arms compared to the inter-arms outside 1\re, which indicates younger stellar age in the arm regions than the inter-arms. These results are in agreement with the observed higher \SigSFR\ within the arm regions of SG1202 (Sec~\ref{subsec:SFR}).
    However, in SG1204, we have difficulty comparing the \dfth\ within the arm regions and the close inter-arms based on the tight spiral features. We carry on our study of \dfth\ by introducing the \dfth\ residual in the next paragraph.
    
	To explore the azimuthal variation of \dfth, we subtract the radial gradient from the original \dfth\ maps (middle panels of Fig~\ref{fig:d4k}) to create the \dfth\ residual maps ($\Delta$\dfth; right panels of Fig~\ref{fig:d4k}). The radial gradient is represented by the weighted average of each 1~kpc elliptical bin (light pink star symbols in the left panels of Fig~\ref{fig:d4k}).
	A positive $\Delta$\dfth\ indicates an older stellar age than the average stellar age at the same galactocentric distance. 
        Both SG1202 and SG1204 show a generally negative $\Delta$\dfth\ (blue spaxels) within the spiral arms.
	The KS test results ($p$-value of 2.0 $\times$ 10$^{-6}$) indicate that the distribution of $\Delta$\dfth\ in the spiral arms of SG1202 is significantly different from the distribution in the inter-arms. 
    Similarly, SG1204 shows a $\Delta$\dfth\ variation between arms and inter-arms, with a $p$-value of 3.6 $\times$ 10$^{-3}$.
    The concentration of negative $\Delta$\dfth\ (younger stellar population) in the arm region of SG1202 and SG1204 is in line with the expectation from the observed higher \SigSFR\ in the arms (Sec~\ref{subsec:SFR}).
    We will further discuss the connection among different tracers in Sec~\ref{sec:discussion}.

\begin{figure*}
	    \centering
	    \includegraphics[height=2.6in]{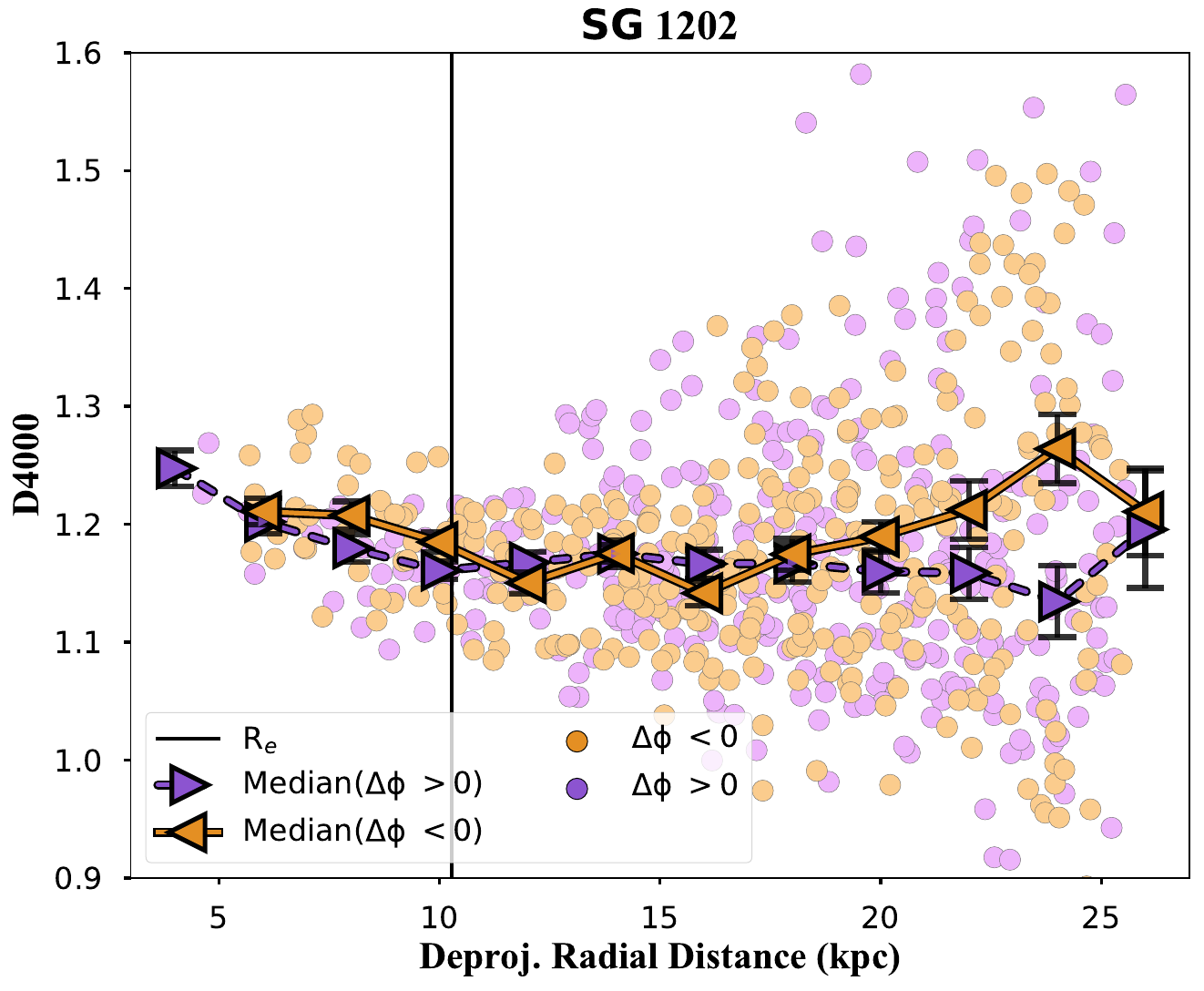}
	    \includegraphics[height=2.6in]{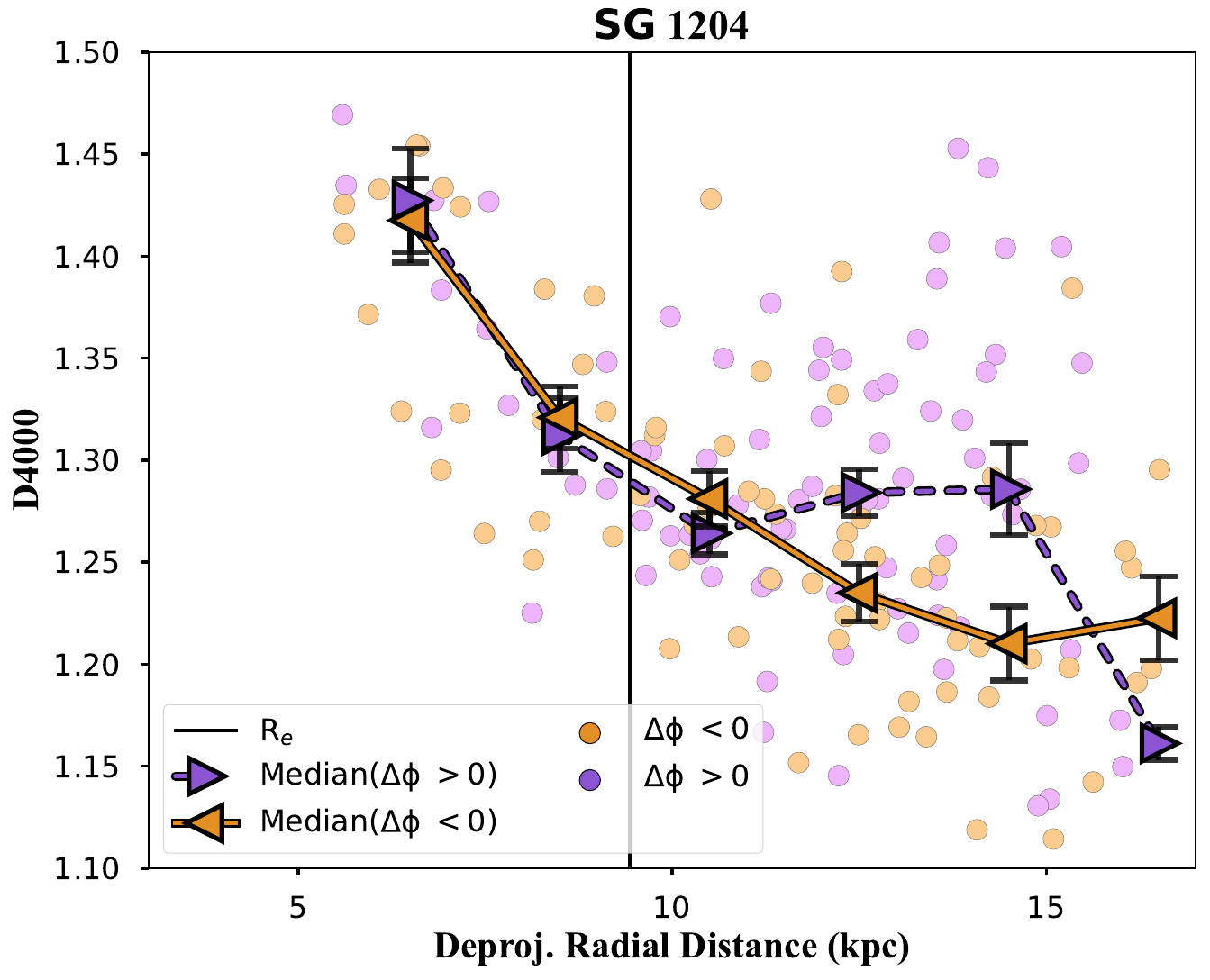}
	    \caption{Similar to Fig~\ref{fig:z_updown} but with \dfth\ on the $y$-axis. We find comparable \dfth\ between \dphi\ $> 0$ (upstream; purple) and \dphi\ $< 0$ (downstream; orange) in both SG1202 and SG1204.} 
	    \label{fig:d4k_updown}
	\end{figure*}

    Taking advantage of the high spatial resolution of MUSE, we compare the \dfth\ in the downstream and upstream across different radii (Fig~\ref{fig:d4k_updown}).  
    We observe no distinguishable \dfth\ offset, within 1$\sigma$ uncertainty, between downstream and upstream in SG1202 from the radial profiles (Fig~\ref{fig:d4k_updown}). 
    The $p$-values from the KS test on the $\Delta$\dfth\ of downstream versus upstream (1.4 $\times$ 10$^{-1}$ for SG1202 and 3.9 $\times$ 10$^{-1}$ for SG1204) indicate the $\Delta$\dfth\ on both sides of the spiral arms are consistent with being drawn from the same distribution.
    The offset of the medians on the outskirt of SG1204 could be attributed to the limited number of available spaxels.

	\begin{table*}
		\newcommand{\tabincell}[2]{\begin{tabular}{@{}#1@{}}#2\end{tabular}}
		\centering
            \small
		\begin{tabular}{ccccc}
			\hline
	    Galaxy    &  gas and stellar properties & Arms vs. Inter-arms & \tabincell{c}{\dphi\ $< 0$ vs. \\ \dphi\ $> 0$} \\
	    \hline
	    \multirow{2}*{SG1202} & \offset & 1.7$\times$10$^{-4}$ & 4.7$\times$10$^{-3}$\\
	    ~ & $\Delta$\dfth & 2.0$\times$10$^{-6}$ & \textbf{1.4$\times$10$^{-1}$} \\
	    \hline
	    \hline
	    \multirow{2}*{SG1204} & \offset & \textbf{6.8$\times$10$^{-2}$} & \textbf{4.1$\times$10$^{-1}$}\\
	    ~ & $\Delta$\dfth & 3.6$\times$10$^{-3}$ & \textbf{3.9$\times$10$^{-1}$}\\
	    \hline
	    \end{tabular}
	    \caption{The $p$-values of KS-test on \offset\ - first and third rows - and $\Delta$\dfth\ - second and fourth rows - in the spiral arms versus inter-arms (column 3), \dphi\ $< 0$ (downstream) versus \dphi\ $> 0$ (upstream) in column 4.  
        The populations whose $p$-values from the KS test are larger than 1\% are highlighted in bold.}
	    \label{tab:ks}
	\end{table*}

    To summarise, the open-armed spiral galaxy SG1202 shows higher \SigSFR\ (left panel of Fig~\ref{fig:sfr_updown}) and lower gas-phase metallicity (left panel of Fig~\ref{fig:z_updown}) in the downstream (\dphi\ $< 0$; orange) than the upstream (\dphi\ $> 0$; purple). As a proxy for the stellar age within SG1202, \dfth\ does not show a distinguishable offset between both sides of the arm regions (left panel of Fig~\ref{fig:d4k_updown}). 
    The tightly wounded spiral galaxy SG1204 presents no significant differences in the \SigSFR\ (right panel of Fig~\ref{fig:sfr_updown}), 12 + log(O/H) (right panel of Fig~\ref{fig:z_updown}) or \dfth\ (right panel of Fig~\ref{fig:d4k_updown}) between the downstream and the upstream.
    Observed differences in the azimuthal distributions of \ha, 12 + log(O/H) and \dfth\ may arise due to the different stellar time scales for these tracers, which will be further discussed in Sec~\ref{sec:discussion}.
	
 \section{Discussion}\label{sec:discussion}
 In this section, we discuss the impacts of the boundary between downstream and upstream regions on the results of the MAGPI survey. We illustrate the differences among various ISM tracers and stellar population indicators. It reminds us of the importance of selecting an appropriate indicator when testing the density wave theory through observations. We also compare our observational results with the theoretical expectations derived from the density wave theory and the dynamical spiral theory. In the last subsection, we address the effects of gas flows, which can complicate our test of the density wave theory. 
 
\subsection{Impacts of the boundary between downstream and upstream regions}
\begin{figure}
	    \centering
	    \includegraphics[height=2.6in]{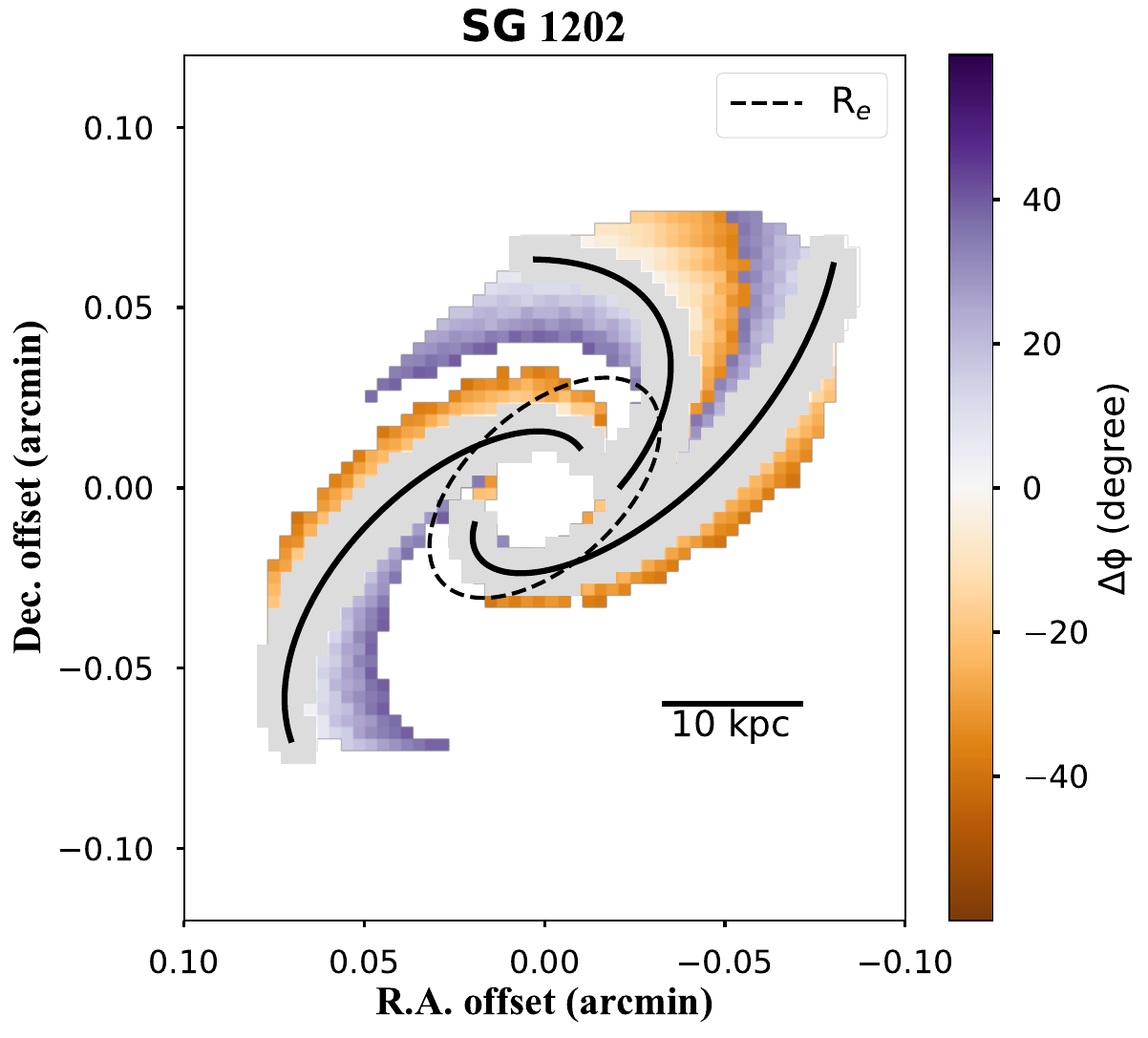}
	    \caption{\dphi\ map after applying the limit of $|$\dphi$|$ < 40$^{\circ}$. The grey regions are defined as the spiral arm regions. The limit of the color bar is matched with Fig~\ref{fig:delta_phi} for better comparison.} 
	    \label{fig:dphi40}
	\end{figure}

\begin{figure*}
	    \centering
	    \includegraphics[height=1.8in]{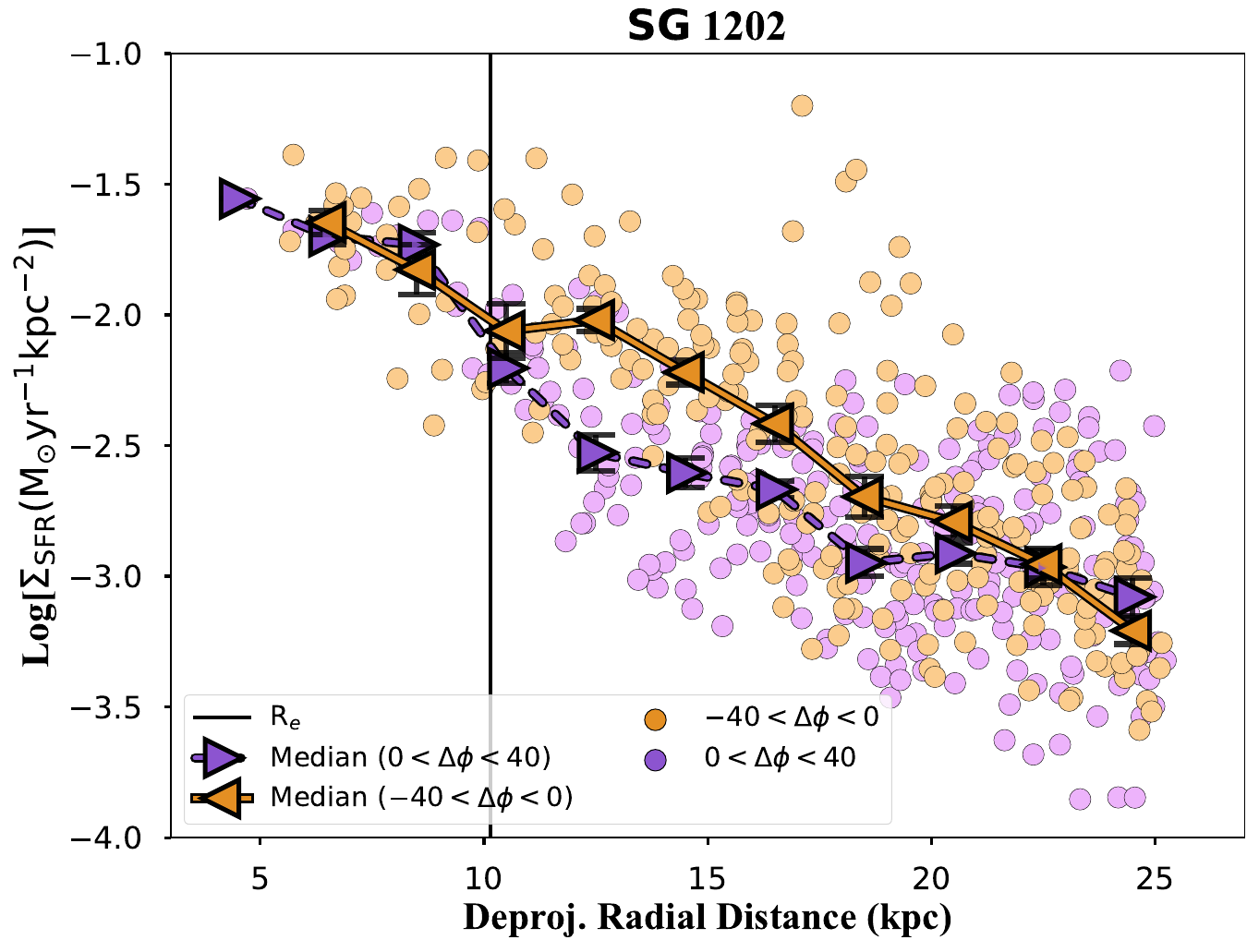}
	    \includegraphics[height=1.8in]{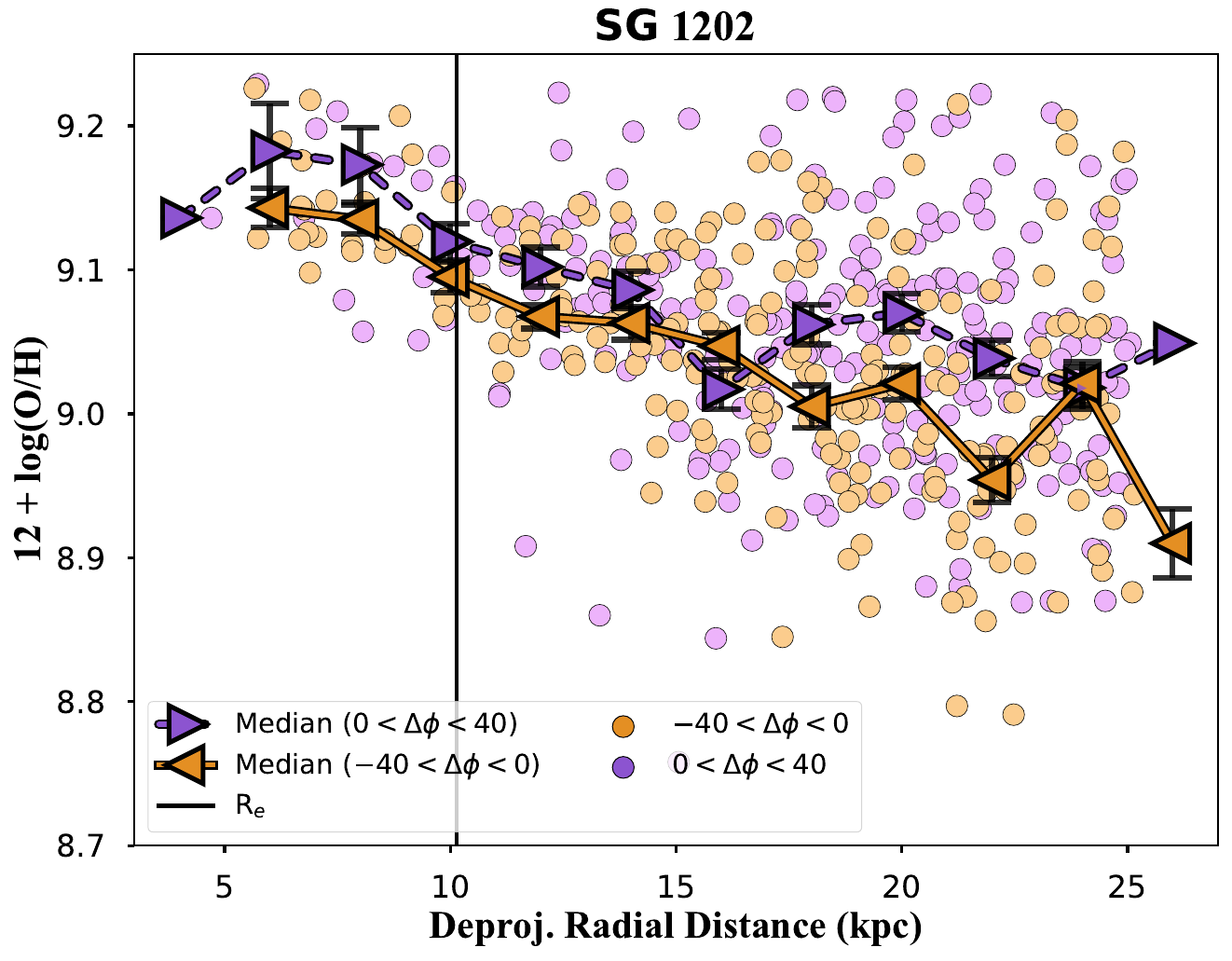}
            \includegraphics[height=1.8in]{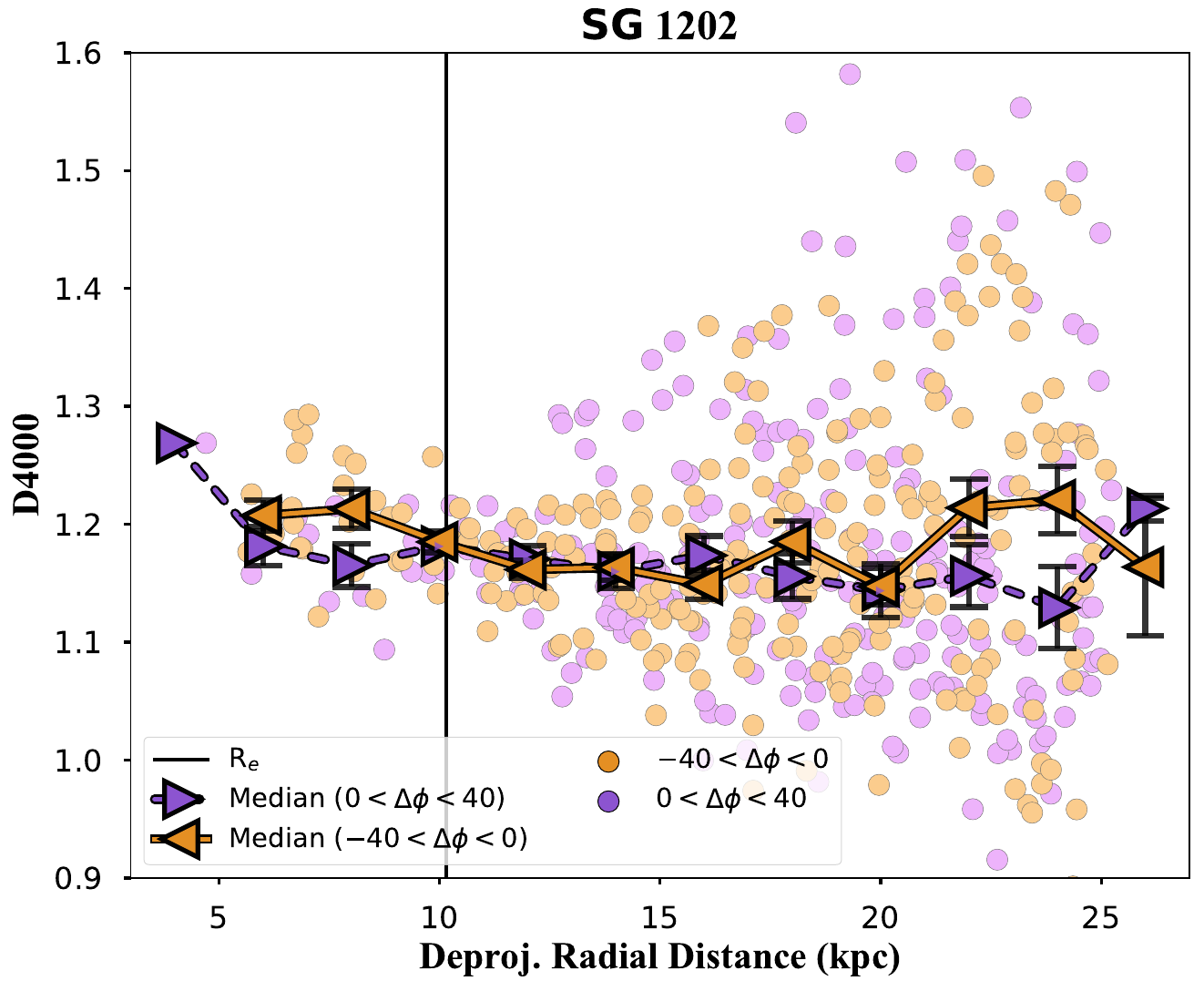}
	    \caption{Radial profiles of \SigSFR, 12+log(O/H), and \dfth, colour-coded by the downstream (orange) and upstream (purple). Only spaxels where $|$\dphi$|$ < 40$^{\circ}$ are considered.} 
	    \label{fig:dphi40_ism}
	\end{figure*}

We find enhanced \SigSFR\ in the downstream (orange) than the upstream (purple) in SG1202 but not in \dfth. 
It is important to note that the connection between downstream and upstream at the high $|$\dphi$|$ end can possibly obscure the signals of offset. 
To address this, we conduct a test with SG1202 by employing a $|$\dphi$|$ limit of < 40$^{\circ}$ (see Fig~\ref{fig:dphi40}) and subsequently re-compare the downstream and upstream regions. 
Similarly, we detect a significant azimuthal variation in \SigSFR\ but not in the 12+log(O/H) or \dfth\ (Fig~\ref{fig:dphi40_ism}).
The $p$-value (1.6 $\times 10^{-3}$) from the KS test suggests the 12+log(O/H) on both sides of the spiral arms are drawn from different distributions. 
On the other hand, the $p$-value (7.3 $\times 10^{-1}$) from the KS test indicates the \dfth\ on both sides of the spiral arms are drawn from the same distribution.
These results are consistent with our main findings, where we do not impose a limit on the $|$\dphi$|$ (see Sec~\ref{sec:analysis}).
Our tests suggest that the linkage between the downstream and the upstream has minimal impact on the azimuthal variation.

 \subsection{Difference in various tracers}\label{sec:3ism}
    The interpretation of the physics driving our observational trends is impacted by the different timescales associated with and traced by the ISM and stellar properties.
    \citet{Poggio_2022} find the progressive disappearance of spiral structures in the global metallicity when they gradually include cooler and older stars. 
    Their result hints at the importance of using tracers sensitive to appropriate time scales when observing spiral structures and material in the inter-arm regions.
    We compare and summarise what the different indicators (SFR, gas-phase metallicity and \dfth) trace and their different timescales in this section.

    Only stars more massive than $\sim$ 20 M$_{\odot}$ produce a measurable ionising photon flux. For recently formed stellar populations born through an instantaneous burst, the ionising photon flux decreases by two orders of magnitude between 5 Myr and 10 Myr after the burst. 
    Thus, the \ha\ emission is an instantaneous tracer \citep[$\lesssim$ 30 Myr;][]{Calzetti_2013} of the youngest, most massive and short-lived rare stars.
    
    Gas-phase metallicity traces metals produced by recent star formation, and it reaches equilibrium within a concise time (a few hundred Myr) compared to the depletion time \citep[][]{Dave_2011, Dave_2012, Lilly_2013, Sharda_2021}. 
    The emission lines we use to measure metallicity are mostly excited by O and B stars, which have formed within the past $\lesssim$ 10 Myr, and the gas-phase metallicity measured in this way can therefore be treated as the current "instantaneous" metallicity of the gas out of which the stars have formed.
     
	The \dfth\ is tracing the stellar population ages while \ha\ and gas-phase metallicity are indicators for gas ISM properties. 
     The drop in intensity at the blue end of the spectrum is a result of (1) a relative lack of young, bright, blue stars in old stellar populations that dominate the light of a galaxy and (2) increased metal absorption in stellar atmospheres of old and metal-rich stellar populations.
     \dfth\ is sensitive to starbursts in the past 1 $\sim$ 2~Gyr \citep[][]{Kauffmann_2003a}, which is tracing a timescale significantly longer than that of \ha\ emission and gas-phase metallicity. The long timescale may allow the latest generation of stars to travel across the disc and reach the next spiral pattern.
	Due to their rarity and short-lived nature, the massive stars make up a small portion of all stars in a galaxy and thus have a small effect on the mean \dfth\ in the IFS data as opposed to low-mass, cooler and longer-lived stars.
     Therefore, the influence of spiral structures on \dfth\ is less significant, compared to \ha\ as an emission line feature.

        We estimate the upper limit of crossing time ($t$) between two spiral arms by assuming the spiral arms have an equal distance to each other at a radius of 2\re, using the following equations.
        \begin{equation}
            D_{\mathrm{arm}} = \frac{4 \pi R_e}{3},
        \end{equation}
        where $D_{\mathrm{arm}}$ is simply the circumference at 2\re\ divided by three (for a three spiral-arm galaxy),
        \begin{equation}
            t = \frac{D_{\mathrm{arm}}}{v_{\mathrm{rot}}}.
        \end{equation}
    
        We adopt the gas rotational velocity ($v_{\mathrm{rot}}$) from Sharma et al. (in prep), which yields the beam-smearing corrected rotational curve based on their 3D modelling method of the ionised gas kinematics. 
        They find that $v_{\mathrm{rot}}$ at 2\re\ is 225~km/s and 305~km/s in SG1202 and SG1204, respectively. 
        The galactocentric distance of 2\re\ is 20~kpc for SG1202 and 18~kpc for SG1204. 
    The estimated crossing time of SG1202 (SG1204) is $\sim$0.18~Gyr ($\sim$0.12~Gyr), which is shorter than the timescale to which \dfth\ is sensitive (1 $\sim$ 2~Gyr). Thus the absent azimuthal variation in the \dfth\ of SG1202 as well as SG1204 is not unreasonable. 
    Both estimated crossing times are longer than the timescale to which \ha\ is sensitive (i.e., $\lesssim$ 30 Myr). Therefore, the observed azimuthal variation in \ha\ of SG1202 is reasonable.
    
    Our work reveals the underlying differences in timescales traced by gas ISM properties (\ha\ and gas-phase metallicity) and stellar populations (\dfth). 
    When investigating the mechanisms driving spiral features, selecting an indicator sensitive to a timescale shorter than the crossing time $t$ is crucial. 
    Our pilot sample suggests that an instantaneous or short-time-scale tracer (e.g., SFR and gas-phase metallicity) is better suited for testing the density wave theory, compared to \dfth.

\subsection{Testing the density wave theory}
    Recently, there has been a growing number of observational tests on the density wave theory and dynamic spiral theory, which are mostly regarding the pattern speed of spiral arms and the differentiation of gas and stellar motion.
    Due to the challenges in measuring the pattern speed, an indirect method to constrain the origin of spiral features is to determine the location of spiral shocks. The density wave theory expects an offset between the spiral shocks and the spiral pattern while there is no observable offset expected by the dynamic spirals.
   In SG1202, we find significantly higher ($\sim$ 0.25 dex) \SigSFR\ in the leading edge of the spiral arms (downstream; orange). The azimuthal variation in \SigSFR\ suggests that the origin and development of the spiral arms in SG1202 can be explained by the density wave theory.

    Based on the density wave theory, \citet{Edmunds_1990} and \citet{Ho_2017} construct chemical evolution models which predict an azimuthal variation in the gas-phase metallicity. NGC1365 \citep{Ho_2017} is observed to show azimuthal variation in the gas-phase metallicity which is consistent with the enhancement effects of spiral arms when material flows across the density waves. 
    However, there are some nearby spiral galaxies showing no azimuthal variation in metallicity \citep{Kreckel_2019}.

    In this work, we do not observe a significant azimuthal variation in metallicity in SG1202, unlike NGC~1365 \citep[$\sim$ 0.2~dex;][]{Ho_2017}. 
    The KS-test (Sec~\ref{subsec:z}) shows that the gas-phase metallicities in the downstream and upstream regions of SG1202 are drawn from different distributions, which is consistent with the chemical evolution models based on the density wave theory.

    \citet{Dobbs_2010} suggest an indirect method to examine the density wave theory through the observed age distribution of stellar clusters. 
    Based on the density wave theory \citep{Lin_1964}, star formation occurs when the gas is compressed and shocked as it periodically passes through the spiral density wave. 
    Due to the differential rotation between material and density wave, the newborn stars move out of the spiral pattern, where the stellar populations inside the co-rotation radius overtake the spirals while the stellar populations outside the co-rotation radius are left behind.
    This is observable as a red spiral arm with an older stellar population that shows a smaller pitch angle than a blue spiral arm with young stellar ages \citep[][]{Martinez-Garcie_2009, Pour-Imani_2016}.
    The density wave theory would thus be observable through an offset between different cluster ages azimuthally across the spiral, whereas dynamic spirals predict no age pattern as material falls into the minimal potential from both sides of the spiral arms, resulting in stars at similar ages lying on both sides of the spirals.
    This method is widely applied to spatially resolved data \citep{Martinez-Garcie_2009,Sanchez-Gil_2011,Shabani_2018,Abdeen_2022}, supporting the density wave theory. On the other hand, there are studies that show no age pattern \citep{Choi_2015, Shabani_2018}, in agreement with dynamic spirals.

    The limited SNR of the continuum in the MAGPI data hinders our ability to determine reliable and resolved stellar ages through SED fitting. 
    In this study, we employ \dfth\ as a stellar age proxy with the necessary spatial resolution.
    In both SG1202 and SG1204, we do not find a statistically significant offset of \dfth\ between the downstream and upstream regions. 
    As detailed in Sec~\ref{sec:3ism}, \dfth\ is not an ideal indicator for testing the density wave theory due to its long timescale (1 $\sim$ 2~Gyr). 
    Thus, it remains inconclusive to compare the observations and theories regarding stellar ages in the context of density wave theory in this work.
        
    Some of our observations provide evidence that challenges the validity of the density wave theory.
    In SG1204, we observe no azimuthal variation in either \SigSFR\ or gas-phase metallicity, which shows a preference for the dynamic spiral theory. 
    It is a caveat that the lack of azimuthal offset results from the combination of i) more pronounced beam-smearing effects than SG1202, and ii) a shorter mixing timescale in SG1204, which features tighter pitch angles.
    We need to expand our study to more MAGPI spiral galaxies to have a better understanding of how pitch angles impact our test of the density wave theory.

\subsection{Radial gas flows complicate the comparison between observations and theories}
    Previous simulations \citep{DiMatteo_2013, Grand_2016, Carr_2022, Orr_2022} have discussed the impact of radial stellar migration and gas flows, induced by asymmetric structures such as spiral arms and stellar bars, on the radial and azimuthal variations of metals.
    In the Feedback In Realistic Environments (FIRE-2) simulation, \citet{Orr_2022} find that the spiral arms act as highways to transfer enriched gas outward, flattening the metallicity radial gradients.
    The FIRE-2 simulated galaxies show no bias of gas-phase metallicity between the arm and inter-arm regions when the sub-grid turbulent diffusion of metals is considered.
    Simulations by \citet{Khoperskov_2023} reveal that the peak of ISM metal-rich material varies depending on the initial radial ISM metallicity gradient.
    Specifically, the metal-rich gas is concentrated as a spiral feature in the trailing side of the stellar spiral arm when a radial gradient is present in the simulation. 
    Their simulation does not see large-scale azimuthal variation in metallicity in a spiral galaxy set with no radial gradient.
    The radial gas flows are important and non-negligible as they potentially impact and complicate the observed metallicity gradient and azimuthal variation. 
    Consequently, they obscure potential observational signatures of density waves and affect our interpretation of the origin of spiral features. 

    We detect a negative gas-phase metallicity radial gradient in both SG1202 ($-0.069\pm0.002$ dex/$R_e$) and SG1204 ($-0.014\pm0.007$ dex/$R_e$), consistent with an inside-out galaxy formation. 
    The gradients we measure for these two galaxies are relatively shallow compared to those typically seen in local spiral galaxies in the CALIFA survey \citep[$-0.1$ dex/$R_e$;][]{SanchezS_2015}, and in the MaNGA survey \citep[$-0.14$ dex/$R_e$;][]{Francesco_2017}.
    This scenario could be indicative of radial gas flows dispersing and flattening the metal distribution.

    The higher gas-phase metallicity observed in the trailing edge of SG1202 (Fig~\ref{fig:z}) is consistent with the simulated galaxy with an initial radial gradient in \citet{Khoperskov_2023} 
    The metallicity distribution in SG1202 could be driven by the density wave theory and/or radial gas flows.
    In SG1204, the azimuthal variation of metallicity is non-observable with a flat metallicity radial gradient.
    This scenario is in line with \citet{Khoperskov_2023} where the presence of radial metallicity gradient is essential for azimuthal variation.
    The dynamic spirals can also result in no azimuthal variation in metallicity.
    Our current observations are insufficient to discern the dominant physical mechanisms driving the spiral features in SG1202 and SG1204.

 \section{Conclusions}\label{sec:conclu}
	We study the two-dimensional distributions of \SigSFR\ (\ha), gas-phase metallicity, and \dfth\ in two three-armed spiral galaxies, SG1202 and SG1204, at $z \sim 0.3$ (3-4~Gyr ago) from the MAGPI survey \citep{Foster_2021}.
    The goal of our study is to investigate the driving physics of spiral arm formation and how the different physical mechanisms impact and vary the ISM and stellar properties  within and up/down stream from the spiral arms.
    Specifically, we compare our findings with the expectations from two spiral arm formation mechanisms: density wave theory and dynamic spirals.
    
    In SG1202, we observe higher \SigSFR\ (Fig~\ref{fig:sfr}), systematically lower metallicity (Fig~\ref{fig:z}), and younger stellar age (Fig~\ref{fig:d4k}) in the spiral arms than the inter-arms.
    In SG1204, we do not find a statistically significant difference in metallicity (Fig~\ref{fig:z}) or \dfth\ (Fig~\ref{fig:d4k}) between the arms and inter-arms. We do, however, recover higher \SigSFR\ (Fig~\ref{fig:sfr}) in the arms of SG1204. These results support the enhancement of star formation within the arm regions.
    The diversity in the metallicity and stellar age azimuthal trends between SG1202 and SG1204 may be a result of a combination of beam-smearing effects and/or shorter mixing timescale, likely due to the tighter pitch angles in SG1204 compared to SG1202.
    
    We find higher \SigSFR\ in the leading edge of spiral arms (\dphi\ $< 0$; downstream) in SG1202 (Fig~\ref{fig:sfr_updown}). This matches the expectation of the density wave theory. 
    The observed higher metallicity in the trailing edge of the spiral arms of SG1202 (Fig~\ref{fig:z_updown}) is consistent with the expectations of radial gas flows driven by spiral arms \citep{Khoperskov_2023}.
    SG1204, on the other hand, does not show a statistically significant offset in the \SigSFR\ (Fig~\ref{fig:sfr_updown}) or metallicity (Fig~\ref{fig:z_updown}) between both sides of the spiral arms.
    Our findings for SG1204 are consistent with the dynamic spiral theory.
    This could also be impacted by radial gas flows, which is consistent with the simulated spiral galaxies with no radial gradients in \citet{Khoperskov_2023}.
    We are however unable to determine the dominant mechanism driving the observations based on the current data.
    
    Neither galaxy shows an azimuthal variation in the \dfth\ (Fig~\ref{fig:d4k_updown}), which may be a consequence of the long timescale ($1 \sim 2$~Gyr) traced by \dfth.
    This reflects the importance of choosing an indicator sensitive to changes in star formation on time scales shorter than the travel time between two spiral arms.
     
	The effects of spiral arms on the ISM are a complex question requiring further high-resolved observational data across cosmic time. This work is a pilot project with two spiral galaxies at $z \sim 0.3$, with a plan to expand the data to the entire MAGPI survey.

\section*{Acknowledgements}
    We thank the anonymous referee for their constructive and detailed comments that improved the quality of this work.
    We thank Paul Torrey for his constructive and enlightening comments and suggestions on this work.
    
    We wish to thank the ESO staff, and in particular the staff at Paranal Observatory, for carrying out the MAGPI observations. MAGPI targets were selected from GAMA. GAMA is a joint European-Australasian project based around a spectroscopic campaign using the Anglo-Australian Telescope. GAMA is funded by the STFC (UK), the ARC (Australia), the AAO, and the participating institutions. GAMA photometry is based on observations made with ESO Telescopes at the La Silla Paranal Observatory under programme ID 179.A-2004, ID 177.A-3016. 
    Parts of this work are supported by the Australian Research Council Centre of Excellence for All Sky Astrophysics in 3 Dimensions (ASTRO 3D), through project number CE170100013.
    KG is supported by the Australian Research Council through the Discovery Early Career Researcher Award (DECRA) Fellowship (project number DE220100766) funded by the Australian Government. 
    PS is supported by Leiden University Oort Fellowship and the IAU Gruber Foundation Fellowship.
    CF is the recipient of an Australian Research Council Future Fellowship (project number FT210100168) funded by the Australian Government. CL, JTM and CF are the recipients of ARC Discovery Project DP210101945.
    GS thanks SARAO postdoctoral fellowship (Grant no.: 97882).
    LMV acknowledges support by the German Academic Scholarship Foundation (Studienstiftung des deutschen Volkes), the Marianne-Plehn-Program of the Elite Network of Bavaria, and the COMPLEX project from the European Research Council (ERC) under the European Union’s Horizon 2020 research and innovation program grant agreement ERC-2019-AdG 882679.
    
    This research has made use of NASA's Astrophysics Data System Bibliographic Services (ADS). 
    This research made use of {\sc astropy},\footnote{\href{http://www.astropy.org}{http://www.astropy.org}} a community-developed core Python package for Astronomy \citep{astropy13, astropy18}. 
    This research has made use of the NASA/IPAC Extragalactic Database (NED) which is operated by the Jet Propulsion Laboratory, California Institute of Technology, under contract with NASA. 
    Parts of the results in this work make use of the colormaps in the {\sc{cmasher}} package \citep{vanderVelden2020}.
	
 \section*{Data Availability}
The data used in this article are available under the MAGPI survey on the ESO public archive (1104.B-0536). The reduced MAGPI MUSE datacubes and emission line data products will be made publicly available in forthcoming papers by Medel et al. (in prep.) and Battisti et al. (in prep), respectively. Additional data generated by the analyses in this work are available upon request to the corresponding author.

\bibliographystyle{mnras}
\bibliography{example} 

\appendix
\section{Identified spiral arms from {\sc sparcfire}}\label{appen:sparc}
\begin{figure}
    \centering
    \includegraphics[width = 0.233\textwidth, height=1.8in]{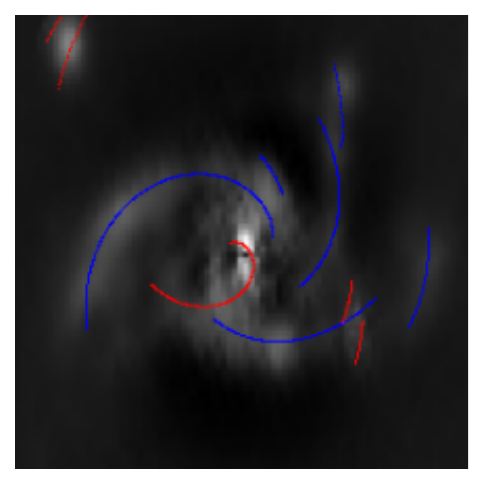}
    \includegraphics[width = 0.233\textwidth, height=1.8in]{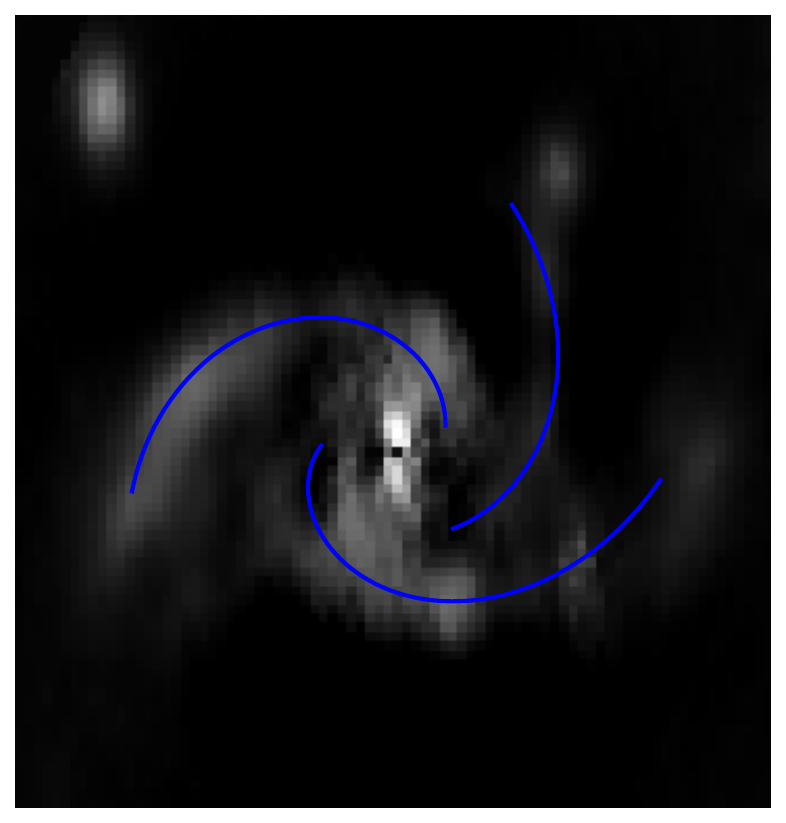}
    
    \includegraphics[width = 0.233\textwidth, height=1.8in]{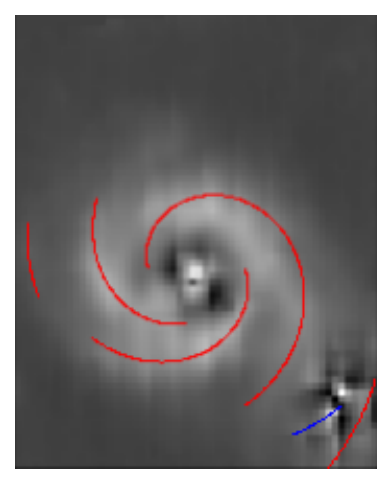}
    \includegraphics[width = 0.233\textwidth, height=1.8in]{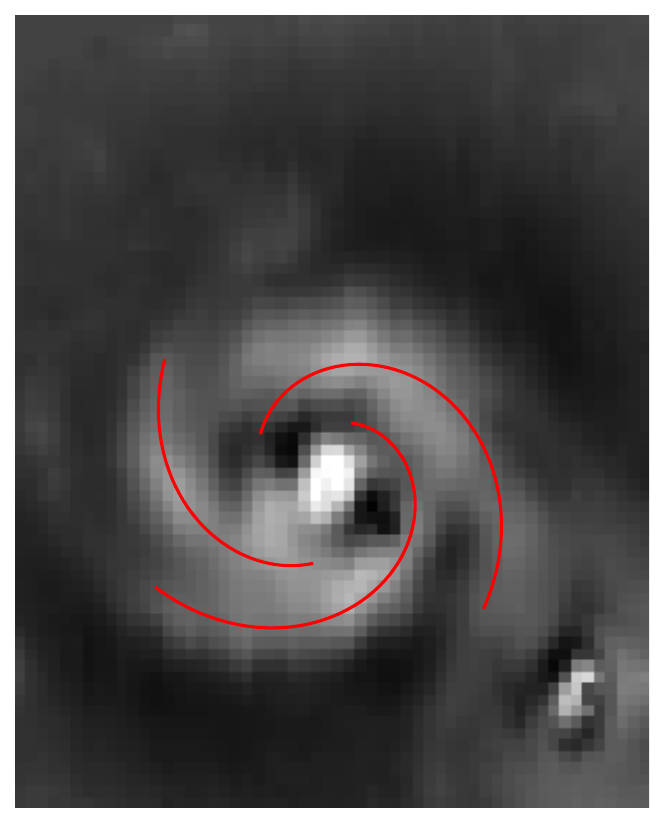}
    \caption{\textbf{Left Column}: The spiral arm ridge initially defined by {\sc{sparcfire}}, overplotted on the deprojected white-light images after subtracted by the Sérsic profile. The red spiral arms are clockwise while the blue ones are counter-clockwise.
    \textbf{Right Column}: The final accepted spiral arm ridge after we fine-tune the starting and ending radii of the spiral arms to improve their alignment with the observed white-light images.
    The top panels show the SG1202 while the bottom panels show the SG1204.}
    \label{fig:sparcfire}
\end{figure}

In Sec~\ref{subsec:define_arm}, we introduce our method of defining the spiral arm ridge lines. The initially defined spiral arms from {\sc{sparcfire}} is overplotted on the deprojected white-light images of our galaxies in the left column of Fig~\ref{fig:sparcfire}. 
We fine-tune the starting and ending radii of the spiral ridge lines to improve their alignment with the observed white-light images, shown in the right column of Fig~\ref{fig:sparcfire}. 
The red spiral arms are clockwise with positive pitch angles while the blue ones are counter-clockwise with negative pitch angles.

\end{CJK} 	
\end{document}